\documentclass[aps,superscriptaddress,nofootinbib,12pt,tightenlines]{revtex4}



\usepackage{natbib}

\usepackage{graphicx}
\usepackage{dcolumn}
\usepackage{amsmath}
\usepackage{color}

\begin{document}


\title{How efficiency shapes market impact}



\author{J. Doyne Farmer}
\affiliation{Institute for New Economic Thinking at the Oxford Martin School and Mathematical Institute, University of Oxford, Eagle House, Walton Well Road, OX2 6ED} 
\affiliation{Santa Fe Institute, 1399 Hyde Park Road, Santa Fe, NM 87501}

\author{Austin Gerig}
\affiliation{CABDyN Complexity Centre, Sa\"{i}d Business School, University of
Oxford, Oxford, United Kingdom}

\author{Fabrizio Lillo}
\affiliation{Scuola Normale Superiore di Pisa, Piazza dei Cavalieri 7, 56126 Pisa, Italy}
\affiliation{Santa Fe Institute, 1399 Hyde Park Road, Santa Fe, NM 87501}
\affiliation{Dipartimento di Fisica e Chimica, viale delle Scienze I-90128, Palermo, Italy}

\author{Henri Waelbroeck}
\affiliation{Portware LLC, 233 Broadway, $24^{th}$ Floor, New York NY 10279}


\begin{abstract}

We develop a theory for the market impact of large trading orders, which we call {\it metaorders} because they are typically split into small pieces and executed incrementally.   Market impact is empirically observed to be a concave function of metaorder size, i.e. the impact per share of  large metaorders is smaller than that of small metaorders.  We formulate a stylized model of an algorithmic execution service and derive a fair pricing condition, which says that the average transaction price of the metaorder is equal to the price after trading is completed.  We show that at equilibrium the distribution of trading volume adjusts to reflect information, and dictates the shape of the impact function.   The resulting theory makes empirically testable predictions for the functional form of both the temporary and permanent components of market impact.   Based on the commonly observed asymptotic distribution for the volume of large trades, it says that market impact should increase asymptotically roughly as the square root of metaorder size, with average permanent impact relaxing to about two thirds of peak impact. 

\end{abstract}


\maketitle

\bigskip
\bigskip
\bigskip
\bigskip
\bigskip
\bigskip


\newpage
\tableofcontents

\section{Introduction}
Market impact is the expected price change conditioned on initiating a trade of a given size and a given sign.  Understanding market impact is important for several reasons.  One motivation is practical: To know whether a trade will be profitable it is essential to be able to estimate transaction costs, and in order to optimize a trading strategy to minimize such costs, it is necessary to understand the functional form of market impact\footnote{
See Bertismas and Lo (\citeyear{Bertismas98}), Almgren and Chriss (\citeyear{Almgren99},\citeyear{Almgren00}), Almgren (\citeyear{Almgren03}), Almgren, Thum, and Hauptmann (\citeyear{Almgren05}), and Obizhaeva and Wang (\citeyear{Obizhaeva05}).}.
Another motivation is ecological: Impact exerts selection pressure against a fund becoming too large, and therefore is potentially important in determining the size distribution of funds\footnote{
See for example Berk and Green (\citeyear{Berk04}).  Schwartzkopf and Farmer (\citeyear{Schwartzkopf10}) have shown that managers of large mutual funds offset increases in market impact by lowering fees, slowing down trading and diversifying assets.}.
Finally, an important motivation is theoretical:  Market impact reflects the shape of excess demand, the understanding of which has been a central problem in economics since the time of Alfred Marshall.

In this paper we present a theory for the market impact of large trading orders that are split into pieces and executed incrementally.  We call these {\it metaorders}\footnote{Other names used in the literature are large trades, packages, or hidden orders.}.  The true size of metaorders is typically not public information, a fact that plays a central role in our theory.  The strategic reasons for incremental execution of metaorders were originally analyzed by Kyle (\citeyear{Kyle85}), who developed a model for an insider trader with monopolistic information about future prices.  Kyle showed that the optimal strategy for such a trader is to break her metaorder into pieces and execute it incrementally at a uniform rate, gradually incorporating her information into the price.   In Kyle's theory the price increases linearly with time as the trading takes place, and all else being equal, the total impact is a linear function of size.  The prediction of linearity is reinforced by Huberman and Stanzl (\citeyear{Huberman04b}) who show that, providing liquidity is constant, to prevent arbitrage permanent impact must be linear.

Real data contradict these predictions:  Metaorders do not show linear impact.  Empirical studies consistently find concave impact, i.e. incremental impact per share decreases with size\footnote{
An early study of metaorders, which links together the individual trades coming from a given client, is Chan and Lakonishok (\citeyear{Chan93}, \citeyear{Chan95}).  Later studies by Torre (\citeyear{Torre97}), Almgren et al. (\citeyear{Almgren05}), Engle et al. (\citeyear{Engle08}), Moro et al. (\citeyear{Moro09}), and Toth et al. (\citeyear{Toth11b}) find concave temporary impacts roughly consistent with a square root functional form.  The functional form of permanent impact is harder to measure and more controversial.  These studies should be distinguished from the large number of studies of the market impact of individual trades or the sum of trades in a given period of time, that do not attempt to link together the individual trades coming from a given client.  See Hasbrouck (\citeyear{Hasbrouck91}), Hausman, Lo and MacKinlay (\citeyear{Hausman92}), Keim and Madhavan (\citeyear{Keim96}), Torre (\citeyear{Torre97}), Kempf and Korn (\citeyear{Kempf99}), Plerou et al. (\citeyear{Plerou02}), Evans and Lyons (\citeyear{Evans02}), Lillo, Farmer, and Mantegna (\citeyear{Lillo03d}), Potters and Bouchaud (\citeyear{Potters03}), Gabaix et al. (\citeyear{Gabaix03,Gabaix06}), Chordia and Subrahmanyam \citeyear{Chordia04}, Farmer, Patelli and Zovko (\citeyear{Farmer05}), Weber and Rosenow (\citeyear{Weber04}), and Hopman (\citeyear{Hopman02}).}.  
It is in principle possible to reconcile the Kyle model with concave dependence on size by making the additional hypothesis that larger metaorders contain less information per share than smaller ones, for example because more informed traders issue smaller metaorders\footnote{
This results in concave dependence on size but preserves linear dependence as a function of time.  Our model in contrast predicts that both size and time follow the same concave functional form.  The empirical results strongly support concave dependence on size, whereas the dependence on time is an open question.}.
A drawback of this hypothesis is that it is neither parsimonious nor easily testable, and as we will argue here, under the assumptions of our model it violates market efficiency.

Huberman and Stanzl are careful to specify that linearity only applies when liquidity is constant.  In fact, liquidity fluctuates by orders of magnitude and has a large effect on price fluctuations\footnote{
Farmer et al. (\citeyear{Farmer04b}) show that fluctuations in instantaneous liquidity can span as much as three orders of magnitude for the same equity in the course of a year; Gillemot et al. (\citeyear{Gillemot06}) show that liquidity fluctuations dominate volume fluctuations in driving clustered volatility.  Some of the theoretical consequences of time varying liquidity have been investigated by Challet (\citeyear{Challet07b}) and Gatheral (\citeyear{Gatheral10}).}.
Empirical studies find that order flow is extremely persistent, in the sense that the autocorrelation of order signs is positive and decays very slowly.  No arbitrage arguments imply either fluctuating asymmetric liquidity    
as postulated by Lillo and Farmer (\citeyear{Lillo03c}), or no permanent impact, as discussed by Bouchaud et al. (\citeyear{Bouchaud04})\footnote{
Asymmetric liquidity means that the price response to a buy order differs from the price response to a sell order of the same size.  The persistence of order flow implies predictability of order signs.  If impact has a non-zero permanent component, then if the next order is likely to be a buy, no-arbitrage implies that the price response to a buy order must be smaller than the price response to a sell order.  See also Bouchaud et al. (\citeyear{Bouchaud04b}), Farmer et al. (\citeyear{Farmer06}), Wyart et al. (\citeyear{Wyart06}), and Bouchaud, Farmer and Lillo (\citeyear{Bouchaud08b}).  For a precursor of the theory developed here see the PhD thesis of Gerig (\citeyear{Gerig07}). For an early attempt to derive a theory yielding a square root market impact see Zhang (\citeyear{Zhang99}).  These theories are for the impact of individual transactions, and the answers they give for the time and size dependence of impact are quite different than those we derive here for metaorders that are typically composed of many individual transactions.}.

The central goal of our model is to understand how order splitting affects market impact\footnote{
The persistence of order flow in the London Stock Exchange has been shown to be overwhelmingly due to order splitting rather than herding by Toth et al. (\citeyear{Toth11c}).}.
Whereas Kyle assumed a single, monopolistic informed trader, our informed traders are competitive.   They receive a common information signal and must independently choose the size of the order to submit.  The orders are submitted to an algorithmic execution service that bundles them together as one large metaorder and executes them incrementally.    We show that this leads to a Nash equilibrium in which order sizes are equilibrated with information.  This equilibrium satisfies the condition that the final price after a metaorder is executed equals its average transaction price.  We call this condition \emph{fair pricing}, to emphasize the fact that under this assumption trading a metaorder is a breakeven deal -- neither party makes a profit as a result of trading\footnote{
Neither party makes a profit in a one way trade based on current prices, but for a round trip the aggressive party nonetheless incurs losses equal to the one way permanent impact.}.
Our equilibrium is less general than Kyle's in that it assumes uniform execution, but it is more general in that it allows an arbitrary information distribution.  This is key because, as we show, there is an equilibrium between information and metaorder size, and the metaorder size distribution can be observed directly from empirical data.

For a given information distribution our theory predicts the metaorder size distribution and the average impact as a function of time during execution. Metaorder size and average impact are both observable, and in some datasets it is even possible to know whether or not trades are informed (for example, Gomes and Waelbroeck (\citeyear{Gomes13})).  The fair pricing condition allows us to make several strong predictions based on a simple set of hypotheses.  For a given metaorder size distribution it predicts the average impact as a function of time both during and after execution.  We thus predict the relationship between the functional form of two observable quantities with no a priori relationship, making our theory falsifiable in a strong  sense. It is worth noticing that in a recent paper Bershova and Rakhlin (\citeyear{Bershova13a}) use proprietary data to perform an empirical analysis of a set of large institutional orders executed at Alliance Bernstein's buy-side trading desk and validate the predictions of our theory (more details will be given in the next Sections).

The falsifiability of our model is in contrast to theories that make assumptions about the functional form of utility and/or behavioral or institutional assumptions about the informativeness of trades, which typically leave room for interpretation and require auxiliary assumptions to make empirical tests. 
For example, Gabaix et al. (\citeyear{Gabaix03,Gabaix06}) have also argued that the distribution of trading volume plays a central role in determining impact, and have derived a formula for impact that is concave under some circumstances. However, in contrast to our model, their prediction for market impact depends sensitively on the functional form for risk aversion\footnote{
Gabaix et al. argue that if risk aversion is proportional to $\sigma^\delta$, where $\sigma$ is the standard deviation of profits, the impact will increase with the size $N$ of the metaorder as $N^{\delta/2}$.  Thus the impact is concave if $\delta < 2$, linear if $\delta = 2$ (i.e. if risk is proportional to variance), and convex otherwise.}. 
Our theory, in contrast, is based entirely on market efficiency and does not depend on the functional form of utility.

An interesting alternative to our model, proposed by Toth et al. (\citeyear{Toth11b}), links the shape of the impact to the diffusivity of prices, which gives a linear shape of the latent order book in the vicinity of the spread.  The shape of the impact function depends on model parameters and on the execution speed; in the limit of high participation rates their result is consistent with a square root function for market impact as a function of size.

%
%

Our work here is related to several papers that study market design.  Viswanathan and Wang (\citeyear{Viswanathan02}), Glosten (\citeyear{Glosten03}), and Back and Baruch (\citeyear{Back07}) derive and compare the equilibrium transaction prices of orders submitted to markets with uniform vs. discriminatory pricing.  Depending on the setup of the model, these prices can be different so that investors will prefer one pricing structure to the other and can potentially be ``cream-skimmed" by a competing exchange\footnote{
For example, see Bernhardt, Hughson, and Naganathan (\citeyear{Bernhardt02})}.
The fair pricing condition we introduce here forces the average transaction price of a metaorder (which transacts at discriminatory prices) to be equal to the price that would be set under uniform pricing.  Fair pricing, therefore, means investors have no preference between the two pricing structures.  On the surface, this result is similar to the equivalence of uniform and discriminatory pricing in Back and Baruch (\citeyear{Back07}).  However, in their paper, this equivalence results because orders are always allowed to be split, whereas ours is a true equivalence between the pricing of a split vs. unsplit order.
 
The organization of the paper is the following. In Section II we give a description of the model and discuss its interpretation.  In Section III we develop the consequences of the martingale condition and show how this leads to zero overall profits and asymmetric price responses when order flow is persistent.  In Section IV we show that any Nash equilibrium must satisfy the fair pricing condition.  In Section V we derive in general terms what this implies about market impact.  In Section VI we introduce specific functional forms for the metaorder size distribution and explicitly compute the impact for these cases.  Finally, in Section VII we discuss the empirical implications of the model, discuss how existing empirical literature supports our theory,  and make some concluding remarks.

\section{Model description}

We study a stylized model of an algorithmic trading service combining and executing orders of long-term traders. This can be thought of as a broker-dealer receiving multiple orders on the same security, or as an institutional trading desk of a large asset manager combining the orders from multiple portfolio managers into one large metaorder.  Our goal is to model the price impact of a metaorder during a trading period in which it may or may not be present.  We set up the model in a stylized manner as a game in which trading takes place across multiple periods.   Before the game starts long-term traders receive an information signal; at the end of the trading period final prices, which reflect the signal, are revealed.  While this is somewhat artificial in comparison to a real market (which has no such thing as a ``final" price), the framework is simple enough to allow us to find a solution, and the basic conclusions should apply more broadly.  The structure of the model is in many respects similar to the classic framework of Kyle (\citeyear{Kyle85}), but with several important differences. In our framework the informed trader does not have a monopoly -- they are competitive profit maximizers. Also, the total number of periods is not known, so the market makers must consider at each step the two possibilities that the game will either stop or continue on the next round. A point-by-point comparison with Kyle's model is made in the conclusions.  

\begin{table}[b]
\caption{{\it Agents in the model}.  The long-term investor is the key agent, who must choose an order size $n_k$ at the beginning of the game based on information $\alpha$.  The market makers are competitive profit optimizers who set prices (but don't know $\alpha$).  The day trader and the algorithmic trading firm are mechanical agents; the day trader provides a background of noisy order flow and the algorithmic trading firm bundles up the long-term trades and submits them in lots of equal size.}
\label{modelSummary}
\begin{tabular}{|l|l|l|l|l|}
\tableline
Agent & action & period & information\\
\tableline
long-term investor & submits order $n_k(\alpha)$ to ATF & $0$ & $\alpha$ \\
algorithmic trading firm (ATF) & submits lot from metaorder & $t$ & $N = \sum_{k=1}^{\mathcal{K}} n_k(\alpha)$ \\
day trader & submits order $F(\eta_t)$ & $t$ & $\eta_t$ \\
market maker (MM) & price quote; MM with best quote & $t$ & combined order $s + F(\eta_t)$\\
&  executes transactions at price $\tilde{S}_t$ & &\\
\tableline
\end{tabular}
\end{table}
 
\subsection{Framework}

\begin{figure}[tc]
\begin{center}
\includegraphics[scale=.4]{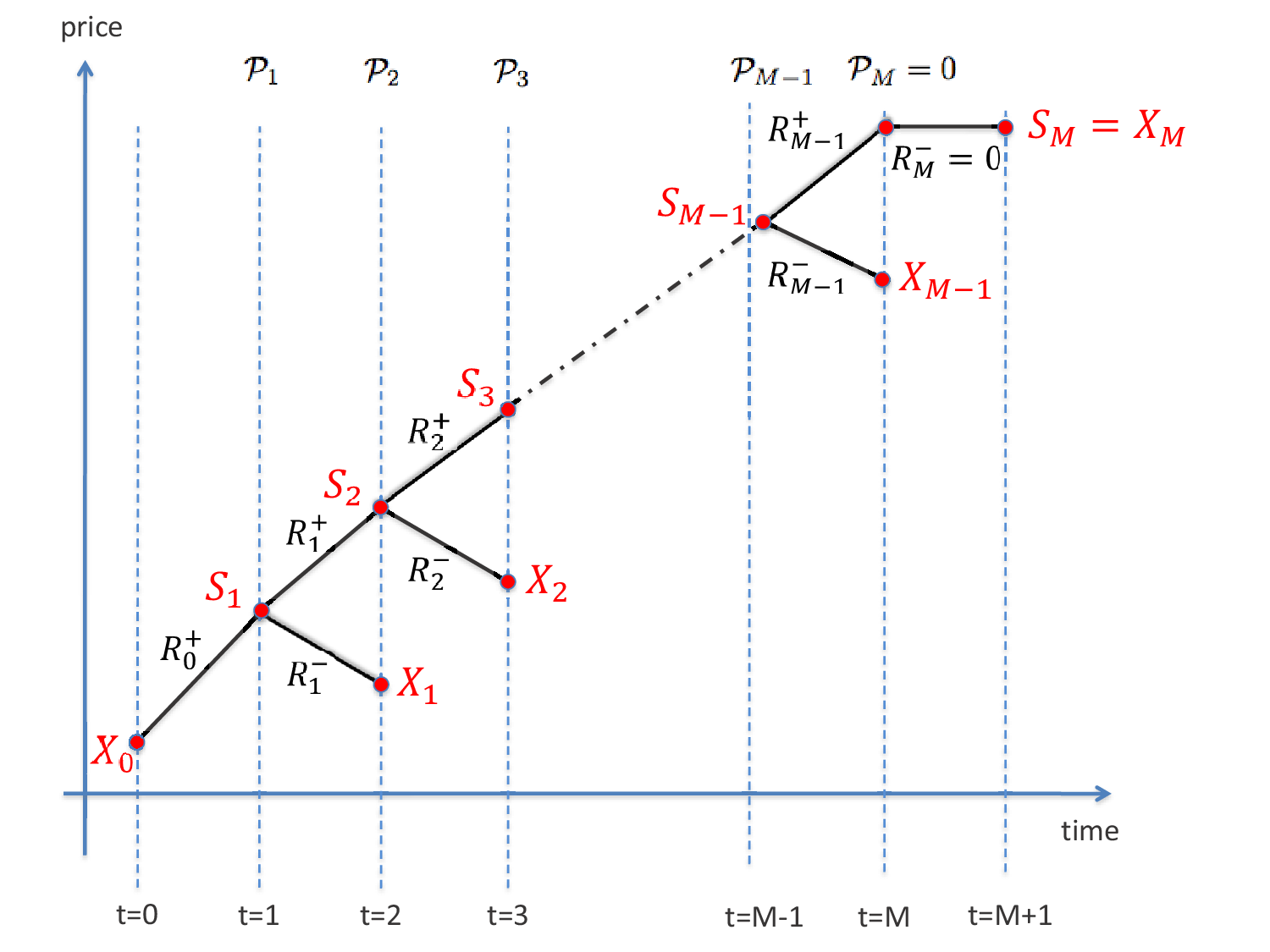}
\caption{The tree of possible price paths for a buy metaorder for different sizes $N$. For this figure we assume the metaorder is present and show only expected price paths, averaged over the day trader's information (which is why the notation does not include tildes).  The price is initially $X_0$; after the first lot is executed it is $S_1 = X_0 + R^+_0$.  If $N = 1$ it is finished and the price reverts to $X_1 = S_1 - R^-_1$, but if $N > 1$ another lot is executed and it rises to $S_2 = S_1 + R^+_1$.  This proceeds similarly until the execution of the metaorder is completed.  At any given point the probability that the metaorder has size $N > t$, i.e. that the order continues, is $\mathcal{P}_t$.  If had we followed a typical price path under circumstances when the day trader's noisy information signal is large, rather than the expected price paths shown here,  the sequence of prices would be a random walk with a time-varying drift caused by the metaorder's impact.}
\label{fig1} 
\end{center}
\end{figure}

The trading of a single asset is organized in a sequence of auctions, which can be regarded as a game.  The auctions take place at times $t = 1, \ldots,T$, where $1 \le T \le M$.  There are three kinds of agents, long-term traders, market makers and a day trader, and an algorithmic trading firm that mechanically bundles orders together.
We use different symbols to indicate different prices. Specifically, $\tilde{S}_t$ is the transaction price at time $t$, while $\tilde{X}_N$ is the final price after the metaorder of size $N$ has finished.  Moreover, as explained in Section II.C, we will average the prices over the different realizations of the day trader's signal.  We indicate with $S_t$ and $X_N$ the averaged transaction and final price, respectively.  The sequence of events can be summarized as follows (see also Figure~\ref{fig1}).  
\begin{description}

\item[$t = 0$:]  The $\mathcal{K}$ long-term traders receive a common information signal $\alpha$, formulate orders of size $n_k(\alpha)$ and submit them to the algorithmic trading firm that bundles them together into a {\it metaorder} of size $\sum_k n_k$.  The algorithmic firm divides the metaorder into equal sized lots (assumed to be of unit size for convenience\footnote{Note that because we measure orders in lots, $n_k$ can be a non-integer fractional size of one lot.})
to be executed incrementally once the game starts.  There may or may not be information:  With probability $\mu$ the signal $\alpha$ is drawn from a distribution $P(\alpha)$, which has nonzero support over a continuous interval $-\alpha_{max} \le \alpha < \alpha_{max}$, where $0 < \alpha_{max} \le \infty$.  With probability $(1 - \mu)$ there is no information, i.e. $\alpha = 0$, which also implies that there will be no metaorder, i.e. $n_k = 0$ for all $k$.   
We assume that the orders of the long-term traders are bounded, i.e. $0 \le n_k (\alpha) \le M_k$, where $M_k=M/\mathcal{K}$ is a large positive integer\footnote{
The imposition of a maximum trade size $M_k$ is a technical detail
to avoid mathematical problems that occur in the limit $M \to \infty$.  This is explained in the Appendix.   For the typical situations we have in mind $M$, $\mathcal{K}$, and $M/\mathcal{K}$ are all large numbers.}.  For simplicity we discuss here the case where $\alpha \ge 0$ but the results apply equally well with obvious modifications when $\alpha \le 0$.

\item[$t = 1, \ldots, T$:]  At each time $t$ a representative {\it day trader} submits a market order of size $F(\eta_t)$, where $\eta_t$ is a random signal, which can be represented as a zero mean IID noise process with an arbitrary distribution $\hat{P}(\eta_t)$, and $F$ is an increasing function whose functional form is not important.  If $\alpha \ne 0$, the algorithmic trading firm also submits a market order of size one.  
The market makers observe the net order flow $s + F(\eta_t)$, where $s=sign(\alpha)$.  The market makers' quotes are formulated independently.  All of the market orders are fully filled by the market maker(s) offering the best prices at the best quote $\tilde{S}_t$ (the meaning of tilde in the notation is explained in Section II.C) .  If the metaorder is present, i.e., if $\alpha \ne 0$, then $T=N$ and the last auction occurs when the metaorder is fully executed, i.e., when 
$$t = N = \sum_k n_k,$$
whereas if $\alpha = 0$, then $T=M$ and the last auction occurs at time\footnote{
We choose the ending time when $\alpha = 0$ to be $M$ because if there were an upper bound $M' < M$ the market maker would know a metaorder was present whenever $t > M'$.}
 $t=M$.
  Given that $\eta_t$ is IID, the typical transaction price sequence $\{\tilde{S}_t\} = \{\tilde{S}_1, \ldots, \tilde{S}_N\} $ will look like a random walk; if there is a metaorder present this induces a time-dependent drift on top of the random walk.  

\item[$t = T+1$:]  The end of the game is announced along with the final price $\tilde{X}_{N}$ (again, the meaning of tilde in the notation is explained in Section II.C). Since the long-term traders' information signals are independent of those of the day trader, information is additive and the final price is 
\begin{equation}
\tilde{X}_{N} \equiv X_0 + \alpha + \sum_{t=1}^N \eta_t,
\end{equation}
where $X_0$ is the initial price.  Note that if no metaorder exists, i.e., $N=0$, then $\tilde{X}_0=X_0+\sum_{t=1}^M \eta_t$.  
\end{description}
We are not concerned here with the question of whether or not the algorithmic execution service's strategy of splitting the order into equal pieces is optimal.  Our goal is instead to take this as given and to derive the implications for price impact. 


\subsection{Further comments about the agents}
We now make some more comments about each of the agents.  We assign a sharp division of labor -- each agent plays an idealized role.  In a more realistic setting, for example, market makers may also be somewhat informed about directional signals.  We do not believe this alters the basic conclusions.

{\bf Long-term traders}.  Only the long-term traders receive the information signal $\alpha$.  For mathematical convenience but without loss of generality we assume that the number of long-term traders is equal to $\mathcal{K}$, where $\mathcal{K}$ is a large number.  The long-term traders are rational:  they know that others have received the same signal, but each long-term trader decides $n_k(\alpha)$ independently.  The decision as to what order size to submit based on the information $\alpha$ is the key decision in the game; this is done only once, in period zero\footnote{It is interesting to note that in a recent paper using proprietary data, Bershova and Rakhlin (\citeyear{Bershova13b}) show that at any given time long-term investors mostly trade on one side of the market, i.e. at any given time they are either buyers or sellers, which gives support to our assumption of common information.}.

{\bf Day traders} provide a noisy background of uncertain order flow that makes it impossible for the market makers to know with certainty whether there is a metaorder present.  The day traders are treated as a single representative agent who receives a private information signal $\eta_t$ at each period $t$ of the game, which is independent of $\alpha$.  The day trader's decision is mechanical; at each period he submits a market order (either to buy or sell) of size $F(\eta_t)$.  The day trader is trading on real information, but nonetheless plays a role in our model similar to that of noise traders in many other models.
The key point is that the day traders do not engage in order splitting, i.e. they trade on the information they receive in given time step only in that time step.  There is no restriction on the size of $\eta_t$, and in particular $\eta_t$ can be large and of the opposite sign of $\alpha$, so that the combined order flow $s + F(\eta_t)$ can be of either sign, even if the steady flow from the long-term traders imparts a bias.  

{\bf Market makers}.  We are not assuming any special institutional privileges, such as those of the specialists in the NYSE; our market makers are simply competitive liquidity providers.  At each time step each market maker observes the combined market order and submits a quote without knowing the quotes of the other market makers.  The combined order is fully executed by the market maker(s) offering the best price.  The market makers are able to take past order flow and prices into account in setting their quotes.

We assume the market makers know the initial price $X_0$, the information distributions $P(\alpha)$ and $\hat{P}(\eta_t)$, the probability $\mu$ that a metaorder is present, and the function $F$ relating the day trader's information to their order size.  During the course of the game they can update their prior $\mu$ to make a time dependent estimate $\mu'_t$.  However they do not know the information signals $\alpha$ or $\eta_t$, and thus they do not know how much of the order flow to ascribe to the long-term trader vs. the day trader.   Thus they do not know with certainty whether a metaorder is present, and if it is, they do not know its size.   

\subsection{Averaging}
We are interested in computing statistical averages.  $\tilde{S}_t$ denotes a specific realization of transaction prices, whereas $S_t$ denotes an average price over the day trader's signals $\eta_i, \forall i\le t$.  Likewise, $\tilde{X}_{N}$ denotes a specific realization of the final price whereas $X_{N}$ denotes an average final price over the day trader's signals $\eta_i, \forall i$.  We use the notation $\hat{E}$ to represent an average over $\eta$,
\begin{equation}
\hat{E}[x(\eta)] = \int x(\eta) \hat{P}(\eta) d\eta.
\end{equation} 
Therefore, $S_t (\alpha) = \hat{E}[ \tilde{S}_t | \alpha]$ and $X(\alpha) = \hat{E}[ \tilde{X} | \alpha ] = X_0 + \alpha$.  The goal of the paper is to compute the {\it average immediate impact} ${\cal I}_t \equiv S_t - X_0$ and the {\it average permanent impact} $I_N \equiv X_{N} - X_0$ of the metaorder.  The corresponding incremental average impacts are $R^+_t = S_{t+1} - S_t$ and $R^-_t = S_t - X_{t}$.  As we will show, at the equilibrium the average impacts do not depend on $\mu$ or $\hat{P}(\eta)$.  However they do depend on the equilibrium distribution of metaorder lengths $p_N$, which is a key quantity that we compute.

In addition to taking expectations over the day trader's random signal $\eta$, which we denote by $\hat{E}$, we must compute expectations over metaorder sizes.  The crux of our argument hinges around the market makers' ignorance of $\alpha$; when $\alpha \ne 0$ this translates into uncertainty about metaorder size.  We will use the notation $E_t$ to represent an average over all metaorders of size $N\ge t$, and as described above, $\hat{E}$ for averages over $\eta_t$.  For a generic function $f_N$ the average over metaorder sizes is
\begin{equation}
E_t[f_N] = \frac{\sum_{N=t}^M p_N f_N}{\sum_{N=t}^M p_{N}}= \frac{\sum_{i=0}^{M-t} p_{t+i} f_{t+i}}{\sum_{i=0}^{M-t} p_{t+i}},
\label{averages}
\end{equation} 
where in the last term we made the substitution $N=t+i$. 

Our main result is to derive the equilibrium relationship between the metaorder size distribution $p_N$ and the average immediate and permanent impacts ${\cal I}_t$ and $I_N$.  

\subsection{Uncertainty of metaorder size and persistence of order flow}


Assuming that a metaorder is present, the likelihood that it will persist depends on the distribution $p_N$ and the number of executions $t$ that it has already experienced.  Let ${\cal P}_t$ be the probability that the metaorder will continue given that it is still active at timestep $t$,
i.e.
\begin{equation}
{\cal P}_t = \frac{\sum_{i=t+1}^{M} p_{i}}{\sum_{i=t}^{M} p_{i}}.
\label{calpdef}
\end{equation}
This makes precise how order splitting can make order flow positively autocorrelated.  In particular, if $p_N$ has tails heavier than an exponential $\mathcal{P}_t$ will increase with time and induce persistence in order flow.  

\subsection{Assumption of known starting and stopping times}  The assumption that the starting and stopping times are known to the participants is similar to the one made in the Kyle model (\citeyear{Kyle85}).  
For long meta-orders and sufficiently high participation rates the starting time can be inferred from the imbalance in order flow\footnote{
The starting time can be inferred by treating order signs as a binomial random process.   The imbalance required to reject the null hypothesis of unbiased order flow is $q \sqrt{t}$, where $q$ is the desired number of standard deviations of statistical significance.  The  accumulated imbalance after $t$ steps is $zt$, where $z$ is the participation rate.  Equating these gives $t = (q/z)^2$.  Hiding an order of size $N$ requires $z < q/\sqrt{N}$.  Thus larger metaorders need to be executed more slowly to avoid detection.  Since the time needed to complete execution is inversely proportional $z$, for a large metaorder this can become prohibitive -- it is impossible to escape detection.}.
Assuming an average participation rate of $20\%$ (i.e. four times as much volume for day traders as from long-term traders) and two standard deviations to reject the null hypothesis of balanced order flow, this means that a metaorder can be detected after about $20$ steps.  Large metaorders are frequently executed in many  small increments\footnote{
Metaorders can be extremely large.  For example, the New York Times recently reported that Warren Buffet took 8 months to buy a $5.5\%$ share of IBM.}.
If there are $1000$ increments the error for inferring the starting time is $2\%$; similarly $10,000$ increments corresponds to an error of $0.2\%$.  Thus for large metaorders it is not unrealistic to imagine that market makers can infer their presence and estimate their starting times and stopping times.  Note, however, that we allow for the possibility that there is no metaorder present at all, and that the probability that the order is not present can be arbitrarily large.  Nonetheless, when $N$ is large the market maker can infer the presence of the order, and the relevant starting and stopping times, with considerable accuracy.

In a more realistic setting there may be many metaorders active at the same time.  Here we assume that we focus attention on a single metaorder, i.e. we are interested in the outcome of an event study in which one examines price changes conditional on the presence of a single metaorder.  As long as arrival of meta-orders are uncorrelated this should be valid.  

\section{Martingale condition}

In this section we introduce a martingale condition and discuss its implications for liquidity and overall profitability.  

\subsection{Derivation of martingale condition}

Market makers must set prices given only past and present order flow information, as well as past prices, without knowing whether the order flow originated from a metaorder or from day traders. Their decision function for setting prices is of the form
\[\tilde{S}_t = f(s + F(\eta_t), s + F(\eta_{t-1}), \ldots, s + F(\eta_1)).
\]
As we show below, we are able to finesse the difficult problem of computing this decision function by imposing a martingale condition and averaging, which is sufficient for the main goal of this paper of deriving the equilibrium between the impacts and $p_N$.

We assume that transaction prices are a martingale, so that the current transaction price $\tilde{S}_t$ is equal to the expected future price.   Define an indicator variable $m$ where $m = 1$ if the metaorder is present and $m = 0$ if it is absent.  Recall that $\mu'_t$ is the market maker's best estimate of the probability that the metaorder is present.  For the price in the next period it is necessary to average over three possibilities\footnote{In these equations, the average is taken over every noise term $\eta_i$.  However, because we condition on $\tilde{S}_t$, the operater $\hat{E}$ does not remove the tilde for this variable.}: 
\begin{enumerate}
\item 
With probability $\mu'_t$ the metaorder is present and with probability $\mathcal{P}_t$ trading continues.   In this case, the expected price in the next period is $\hat{E}[ \tilde{S}_{t + 1} | \tilde{S}_t, m=1] = \tilde{S}_t + R^+_t$.
\item
With probability $\mu'_t$ the metaorder is present and with probability $1 - \mathcal{P}_t$ time $t$ is the last trading period.  In this case, the expected price in the next period is $\hat{E}[ \tilde{X}_{t + 1} | \tilde{S}_t, m =1] = \tilde{S}_t - R^-_t$.
\item
With probability $1 - \mu'_t$ the metaorder is not present. Since the day trader's information $\eta_t$ is zero mean and IID, the expected average transaction price on the next time step conditioned on the current transaction price must satisfy $\hat{E}_{t}[ \tilde{S}_{t + 1} | \tilde{S}_t, m =0] = \tilde{S}_t$, i.e. the average transaction price is unchanged. 
\end{enumerate}

Thus the martingale condition can be written
\begin{eqnarray*}
\nonumber
\tilde{S}_t & = & \mu'_t \left( \mathcal{P}_t E[ \tilde{S}_{t + 1} | \tilde{S}_t, m =1] + (1 - \mathcal{P}_t)E[ \tilde{X}_{t} | \tilde{S}_t, m =1] \right)\\
& & + (1 - \mu'_t) E[ \tilde{S}_{t + 1} | \tilde{S}_t , m =0 ].
\end{eqnarray*}
Substituting for the average next price conditioned on the current price for cases 1-3 gives:
\begin{equation}
\tilde{S}_t = (1 - \mu'_t)\tilde{S}_t + \mu'_t \left(\mathcal{P}_t (\tilde{S}_{t} + R^+_t )+ (1 - \mathcal{P}_t) (\tilde{S}_{t} - R^-_t) \right).
\label{martin1}
\end{equation}
The current transaction price $\tilde{S}_t$ and $\mu'$ both cancel and this reduces to
\begin{equation}
 {\cal P}_t R^+_t - (1 - {\cal P}_t) R^-_t = 0.
 \label{shortterm}
\end{equation}
Equation~(\ref{shortterm}) holds for $t=1,2,..,M-1$. If the metaorder has maximal length at the end of the $M$th interval by definition ${\cal P}_M=0$, which implies that $S_M=X_{M}$.

Let us pause for a moment to digest this result.   We started with a martingale condition for realized prices, including fluctuations caused by day traders, and then reduced it to a martingale condition for the average impact due to the presence of a metaorder.  The reduced martingale no longer depends on $\mu'_t$, $\hat{p}_N$, or  the day trader's information.   The fact that we assume a martingale for realized prices implies that arbitrage of the impact is impossible.  

The ability to average away the day traders is a consequence of our assumption that $\alpha$ and $\eta_t$ are independent, which implies additivity of information.  This separates the problem of the metaorder's impact from that of the day trader's impact -- the metaorder's impact effectively rides on top of the day trader's impact.  As we will see, the virtue of this approach is that it allows us to infer quite a lot without needing to solve for the market makers' optimal quote setting function $f$.

\subsection{Asymmetric price response\label{arbefficiency}}

Equation~(\ref{shortterm}) can be trivially rewritten in the form
\begin{equation}
\frac{R^+_t }{R^-_t} = \frac{1 - {\cal P}_t}{{\cal P}_t},
\label{returnRatio}
\end{equation}
where $R^+_t = S_{t+1} - S_t$ and $R^-_t = S_t - X_{t}$.  
Thus the martingale condition fixes the ratio of the price responses $R^-_t$ and $R^+_t$, but does not fix their scale.  If ${\cal P}_t$ is large, corresponding to a metaorder that is likely to continue, then $R^+_t/R^-_t$ small.  This means that  the price response if the order continues is much less it is than if it stops. 

To complete the calculation we need another condition to set the scale of the price responses $R^-_t$ and $R^+_t$, which may change as $t$ varies\footnote{
Various authors have used alternative conditions.  For example, Gerig (\citeyear{Gerig07}) and Gerig et al.~(\citeyear{Gerig11}) use a symmetry condition, which can be derived from assumptions of linearity.  The fair pricing condition that we derive here has the advantage that it can be justified based on equilibrium arguments.}.
Such a condition is introduced in Section~\ref{sizeIndependenceSec}.   Even without such a condition, one can already see intuitively that ``all else equal", for a heavy-tailed metaorder distribution $p_N$, the impact will be concave.  (Recall that heavy tails in $p_N$ imply that $\mathcal{P}_t$ increases with $t$). 

\subsection{Zero overall profits}

The martingale condition implies that the market makers break even overall, i.e. that their total profits summing over metaorders of all sizes is zero.  This is stated more precisely in proposition one.

{\bf Proposition 1.} The average prices for each transaction are $\{ S_t \}$, with $t = 1, \ldots, N$.  At the final step the valuation is $N X_{N}$.  The martingale condition implies zero overall profits, i.e.
\begin{equation}
\Pi = E_1[N\pi_N]\equiv\sum_{N=1}^M  p_N N \pi_N =0,
\label{breakevenonaverage}
\end{equation}
where
\begin{equation}
\pi_N \equiv \frac{1}{N} \sum_{t=1}^N S_t - X_{N}
\label{immediateProfit}
\end{equation}
is the profit per lot transacted.   
The proof of proposition 1 is given in Appendix A.  The phrase ``overall profits" emphasizes that the martingale condition only implies zero profits when averaged over metaorders of all sizes.  It allows for the possibility that the market makers may make profits on metaorders in a given size range, as long as they take corresponding losses in other size ranges.

Surprisingly, Proposition 1 is not necessarily true when $M$ is infinite.  The basic problem is similar to the St. Petersburg paradox:  As the metaorder size becomes infinite it is possible to have infinitely rare but infinitely large losses.  The conditions under which this holds are more complicated, as discussed in Appendix A.
  
\section{Fair pricing\label{sizeIndependenceSec}}

The martingale condition derived in the previous section sets only the ratio $R^+_t/R^-_t$ at each step $t$ (see Eq. \ref{returnRatio}). We therefore need another condition to derive the values of $R^+_t$ and of $R^-_t$ and thus to obtain the expression for the immediate and the permanent impact. The condition we derive here is the  {\it fair pricing condition}, which states that for any $N$
\begin{equation}   
\pi_N = \frac{1}{N} \sum_{t=1}^N S_t - X_{N} = 0.
 \label{fairPricing}
 \end{equation}
Under fair pricing the average execution price is equal to the final price.  We call this fair pricing for the obvious reason that both parties would naturally regard this as ``fair".  Fair pricing implies that the market makers break even on metaorders of any size, as opposed to the martingale condition, which only implies they break even when averaging over metaorders of all sizes. 

We now derive the fair pricing condition by showing that it is a Nash equilibrium\footnote{
Subsequent to our work here, 
Donier (\citeyear{Donier12}) postulates a very similar condition based on perfect competition between market makers.  Recently Bershova and Rakhlin (\citeyear{Bershova13a}) showed that the fair pricing condition and its consequences on market impact are rather well empirically verified in a set of real metaorders.}. 
We prove the following:

{\bf Proposition 2.}  {\it If the immediate impact $\mathcal{I}_t = S_t- X_0$ has a second derivative bounded below zero\footnote{
Note that this is sufficient for concavity.},
in the limit where the number of informed traders $\mathcal{K} \to \infty$, any Nash equilibrium must satisfy the fair pricing condition $\pi_N = 0$ for $1 < N < M$.  On average market makers profit from orders of length one and take (equal and opposite) losses from orders of length $M$.} 

This result is driven by competition between informed traders.  All informed traders receive the same information signal $\alpha$.  The strategy of informed trader $k$ is the choice of the order size $n_k(\alpha)$.  The orders are then bundled together to determine the combined metaorder size $N = \sum_{k=1}^\mathcal{K} n_k(\alpha)$.  The decision of each informed trader is made without knowing the decisions of others.

The derivation has two steps: first we examine the case $\pi_N \ne 0$ for $1 < N < M$, and show that if others hold their strategies constant, providing the impact is concave and $\mathcal{K}$ is sufficiently large, traders can increase profits by changing strategy.  Secondly we show that if $\pi_N = 0$ there is no incentive to change strategy.  Then we return to examine the cases $N = 1$ and $N = M$, which must be treated separately.  The derivation is given in the Appendix.




In contrast to the martingale condition, which only implies that immediate profits are zero when averaged over size, fair pricing means that they are identically zero for every size.  It implies that no one pays any costs or makes any profits simply by trading in any particular size range.

Since informed traders must formulate order sizes knowing only $\alpha$, the distribution of information $p(\alpha)$ implies the distribution of metaorder size $p_N$.   We will use $p_N$ as a proxy for $p(\alpha)$, which is the key fact allowing us to state our results in terms of the observable quantity $p_N$ rather than $p(\alpha)$, which is much more difficult to observe.

\section{General expressions for impact \label{impactSolution}}
 
In this section we assume that the martingale condition holds for all $N$ and the fair pricing condition holds for $ 1 < N < M$.  This allows us to derive both the immediate impact ${\cal I}_N$ and the permanent impact $I_N$ for any given metaorder size distribution $p_N$.   We later argue that for realistic situations the metaorder size distribution gives rise to a concave impact function, consistent with the Nash equilibrium.

The martingale condition (Eq. \ref{shortterm}) and the fair pricing condition (Eq. \ref{fairPricing}) define a system of linear equations for $S_t$ and $X_t$ at each value of $t$, which we can alternatively express in terms of the price differences $R^+_t = S_{t+1} - S_t$ and $R^-_t = S_t - X_{t}$, where $t = 1, \ldots, M$.  The martingale condition holds for $t = 1,..., M$ and the fair pricing condition holds for $t = 2, \ldots, M -1$.  There are thus $2M-2$ homogeneous linear equations with $2M - 1$ unknowns\footnote{
 The price $S_{M+1}$ does not exist, so $R^+_M$ is not needed.  This reduces the number of unknowns by one.}.
Because the number of unknowns is one greater than the number of conditions there is necessarily an undetermined constant, which we choose to be $R^+_1$.   
 
{\bf Proposition 3.}~{\it The system of martingale conditions (Eq. \ref{shortterm}) and fair pricing conditions (Eq. \ref{fairPricing}) has solution}
\begin{eqnarray}
R^+_t=\frac{1}{t}\frac{p_t}{\sum_{i=t+1}^M p_i}\frac{1-p_1}{\sum_{i=t}^M p_i}R^+_1~~~~~t=2,3,...,M-1\label{solution}\\
R^-_t=\frac{{\cal P}_t}{1-{\cal P}_t} R^+_t~~~~~t=1,2,...,M-1
\label{solution2}
\end{eqnarray}
The proof is given in Appendix A.

An important property of the solution is the equivalence of the impact $\mathcal{I}_t$ as a function of either time or size.  This is in contrast to the prediction of an ``extended" Kyle model under the assumption that traders of different sizes are differently informed, which yields linear impact as a function of time while varying nonlinearly with size.


\subsection{General solution for immediate impact}

Summing Eq. \ref{solution} implies that for $N > 2$ the immediate impact is
\begin{equation}
{\cal I}_t = S_t- X_0= R^+_0 + R^+_1 \left( 1 +\sum_{k=2}^{t-1} \frac{1}{k}\frac{p_k}{\sum_{i=k+1}^M p_i}\frac{1-p_1}{\sum_{i=k}^M p_i} \right),
\label{final}
\end{equation}
For $t= 1$ the immediate impact is $\mathcal{I}_1 = S_1 - X_0 = R^+_0$ and for $t=2$ it is $\mathcal{I}_2 = S_2-X_0=R^+_0 + R^+_1$.  (The meaning of the undetermined constants $R^+_1$ and $R^+_0$ is discussed in a moment).

\subsection{General solution for permanent impact}

The permanent impact $X_{N} - X_0$ is easily obtained.  Making some simple algebraic manipulations
\[
X_{N} =X_{N}-S_N+ S_N =S_N -R^-_N=S_N -\frac{{\cal P}_N}{1-{\cal P}_N}R^+_N.
\]
By combining Eqs. (\ref{final}) and (\ref{calpdef}) we get
\begin{equation}
I_N = X_{N}- X_0= R^+_0 + R^+_1 \left(1+\sum_{k=2}^{N-1}  \frac{1}{k}\frac{p_k}{\sum_{i=k+1}^M p_i}\frac{1-p_1}{\sum_{i=k}^M p_i} - \frac{1-p_1}{N  \sum_{i=0}^{M-N} p_{N+i}}\right).
\label{generalPermanentImpact}
\end{equation}

\subsection{Setting the scale}

We have expressed both the permanent and immediate impact  purely in terms of $p_N$ and the undetermined constants $R^+_0$ and $R^+_1$.  The undetermined constants can in principle be fixed based on the information at the equilibrium.  At the equilibrium information signals in the range $\alpha \in [0, \alpha_1]$ will be assigned to metaorders of size one, with an average size $\bar{\alpha}_1$, signals in the range $\alpha \in (\alpha_1, \alpha_2]$ will be assigned to metaorders of size two, with an average size $\bar{\alpha}_2$, and so on.  The scale of the impact is set by the relations
\begin{eqnarray*}
 R^+_0 = I_1 & = & \bar{\alpha}_1 = \int_{0}^{\alpha_1}\alpha p(\alpha) d\alpha\\
R^+_0 + R^+_1\frac{\mathcal{P}_1}{1 - \mathcal{P}_1} = I_2 & = &  \bar{\alpha}_2 =  \int_{\alpha_1}^{\alpha_2}  \alpha p(\alpha) d\alpha
\end{eqnarray*}
We have used the words ``in principle" because, unlike metaorder sizes, information is not easily observed.           

Barring the ability to independently measure information, the constants $R^+_0$ and $R^+_1$ remain undetermined parameters.   $R^+_1 > 0$ plays the important role of setting the scale of the impact.  The constant $R^+_0$, in contrast, is unimportant -- it is simply the impact of the first trade, before the metaorder has been detected.

\section{Dependence on the metaorder size distribution}

We have so far left the metaorder size distribution $p_N$ unspecified.  In this section we compute the impact for the Pareto distribution, 
\begin{equation}
p_N \sim \frac{1}{N^{\beta+1}}.
\label{paretoEq}
\end{equation}
which we argue is well-supported, at least as an approximation for large $N$, by empirical data\footnote{
The notation $f(x) \sim g(x)$ means that there exists a constant $C\ne 0$ such that in the limit $x \to \infty$, $f(x)/g(x) \to C$.  We use it to indicate that this relationship is only valid in the limit of large metaorder size $N$.}.
In Appendix B we consider the stretched exponential distribution, which is not supported by data, but provides a useful point of comparison.  We also consider a lognormal distribution in Appendix B.

\subsection{Empirical evidence supporting the Pareto distribution\label{empiricalVolume}}

There is now considerable accumulated evidence that in the large size limit in most major equity markets the metaorder size $V$  is distributed as $P(V > v) \sim v^{-\beta}$, with $\beta \approx 1.5$.  

\begin{itemize}
\item
{\it Trade size.} In many different equity markets for large trades the volume $V$ has been observed by several groups to be distributed as a power law (Gopikrishnan et al. (\citeyear{Gopikrishnan00}); Gabaix et al. (\citeyear{Gabaix06})\footnote{
The value of $\beta$ is somewhat controversial, however: Eisler and Kertesz (\citeyear{Eisler06}) and Racz et al. (\citeyear{Racz07}) have argued that the correct value of $\beta > 2$.}.
This relationship becomes sharper if only block trades are considered (Lillo et al., \citeyear{Lillo05b}).
\item
{\it Long-memory in order flow.}  The signs of order flow in many equity markets are observed to have long-memory\footnote{
 Long memory has been observed in the Paris Stock Market by Bouchaud et al. (\citeyear{Bouchaud04}), in the London and New York Stock Markets by Lillo and Farmer (\citeyear{Lillo03c}), and in the Spanish Stock Market by Vaglica (\citeyear{Vaglica08}).  We use the term long-memory in its more general sense to mean any process whose autocorrelation function is non-integrable (Beran, \citeyear{Beran94}).  This can include processes with structure breaks, such as that studied by Ding, Engle and Granger (\citeyear{Ding93}).}   This means that the transaction sign autocorrelation function $C(\tau)$  decays in time as $C(\tau) \sim \tau^{-\gamma}$, where $0<\gamma<1$.  Under a simple theory of order splitting the exponent $\beta = \gamma + 1$,  (Lillo et al., \citeyear{Lillo05b}) and Bouchaud, Farmer, and Lillo (\citeyear{Bouchaud08b}). Since empirically it is $\gamma\simeq 0.5$ this implies $\beta\simeq 1.5$.
\item
{\it Reconstruction of large metaorders from brokerage data.}  Vaglica et al. (\citeyear{Vaglica07}) reconstructed metaorders Spanish stock exchange using data with brokerage codes and found that $N$ is distributed as a power law\footnote{
An interesting point that is relevant for the theory developed here is that the power law behavior of metaorder size comes from the heterogeneity of market participants.  Vaglica et al. (\citeyear{Vaglica07}) showed that metaorder size distribution for individual brokerages is roughly a lognormal distribution, and that a power law only emerges when all the brokerages are combined.  Here we have assumed that the long-term traders are homogeneous, and that $p_N$ is determined by information; this suggests that other factors are at play determining $p_N$, and in particular that the heterogeneous size of investors may be an important factor. }
for large $N$ with $\beta \approx 1.7$.

\item 
{\it Direct evidence from proprietary data.} In a recent paper Bershova and Rakhlin (\citeyear{Bershova13a}) use proprietary data to perform an empirical analysis of a set of large institutional metaorders executed at AllianceBernstein's buy-side trading desk in the U.S. equity market.  
As mentioned above, their data are consistent with an asymptotically Pareto distribution with an estimated tail exponent $\beta=1.56$.
\end{itemize}
There is thus good evidence that metaorders have a power law distribution, though more study is of course needed.

\subsection{Market impact for Pareto metaorder size} \label{paretoSec}

In this section we derive the functional form of the impact for Pareto metaorder size. We do this by using Eq.~\ref{final} in the limit as $M \to \infty$, and return in Section~\ref{finiteSize} and in Appendix B to discuss how this is modified when $M$ is finite.

While we only care about the asymptotic form for large $N$, for convenience we assume an exact Pareto distribution for all $N$, i.e.
\begin{equation}
p_N=\frac{1}{\zeta(\beta)}\frac{1}{N^{\beta+1}},~~~~~~~~~~~~~N\ge1
\label{pareto2Eq}
\end{equation}
where the normalization constant $\zeta(\beta)$ is the Riemann zeta function.  For the Pareto distribution the probability ${\cal P}_t$ that an order of size $t$ will continue is
\begin{equation}
{\cal P}_t=\frac{\zeta(1+\beta,t+1)}{\zeta(1+\beta,t)}\simeq \left(\frac{t}{t+1}\right)^\beta\sim 1-\frac{\beta}{t}.
\label{calPApprox}
\end{equation}
where $\zeta(s,a)$ is the generalized Riemann zeta function (also called the Hurwitz zeta function).  The approximations are valid in the large $t$ limit.

\subsubsection{Immediate impact}

The immediate impact can be easily calculated from Eq. (\ref{final}). 
For Pareto distributed metaorder sizes, using Eqs.~(\ref{solution2}) and (\ref{pareto2Eq}), $R^+_t$ is
\begin{equation}
R^+_t=\left(1+\frac{1}{t^{2+\beta}}\frac{\zeta(1+\beta)-1}{\zeta(1+\beta,t)\zeta(1+\beta,t+1)} \right)R^+_1 \sim \frac{1}{t^{2-\beta}}.
\label{paretoImpact}
\end{equation}
Thus the immediate impact ${\cal I}_t = S_t - X_0$ behaves asymptotically for large $t$ as
\begin{eqnarray}
{\cal I}_t \sim \left \{\begin{array}{ll}
                                                   ~~t^{\beta-1}& {\rm{for}} ~\beta\ne 1 \\
                                                   ~~\log(t+1) & {\rm{for}} ~\beta= 1  \\
                                    \end{array} \right. 
                                    \label{temporaryImpact} 
\end{eqnarray}
The exponent $\beta$ has a dramatic effect on the shape of the impact. For Lorenzian distributed metaorder size ($\beta=1$) the impact is logarithmic, for $\beta=1.5$ it increases as a square root, for $\beta = 2$ it is linear, and for $\beta>2$ it is superlinear.   Thus as we vary $\beta$ the impact goes from concave to convex, with $\beta = 2$ as the borderline case\footnote{
The reason $\beta = 2$ is special is that for $\beta < 2$ the second moment of the Pareto distribution is undefined.   Under the theory of Lillo et al. (\citeyear{Lillo05b}), long-memory requires $\beta < 2$.}.
\begin{figure}[ptb]
\begin{center}
\includegraphics[scale=0.4]{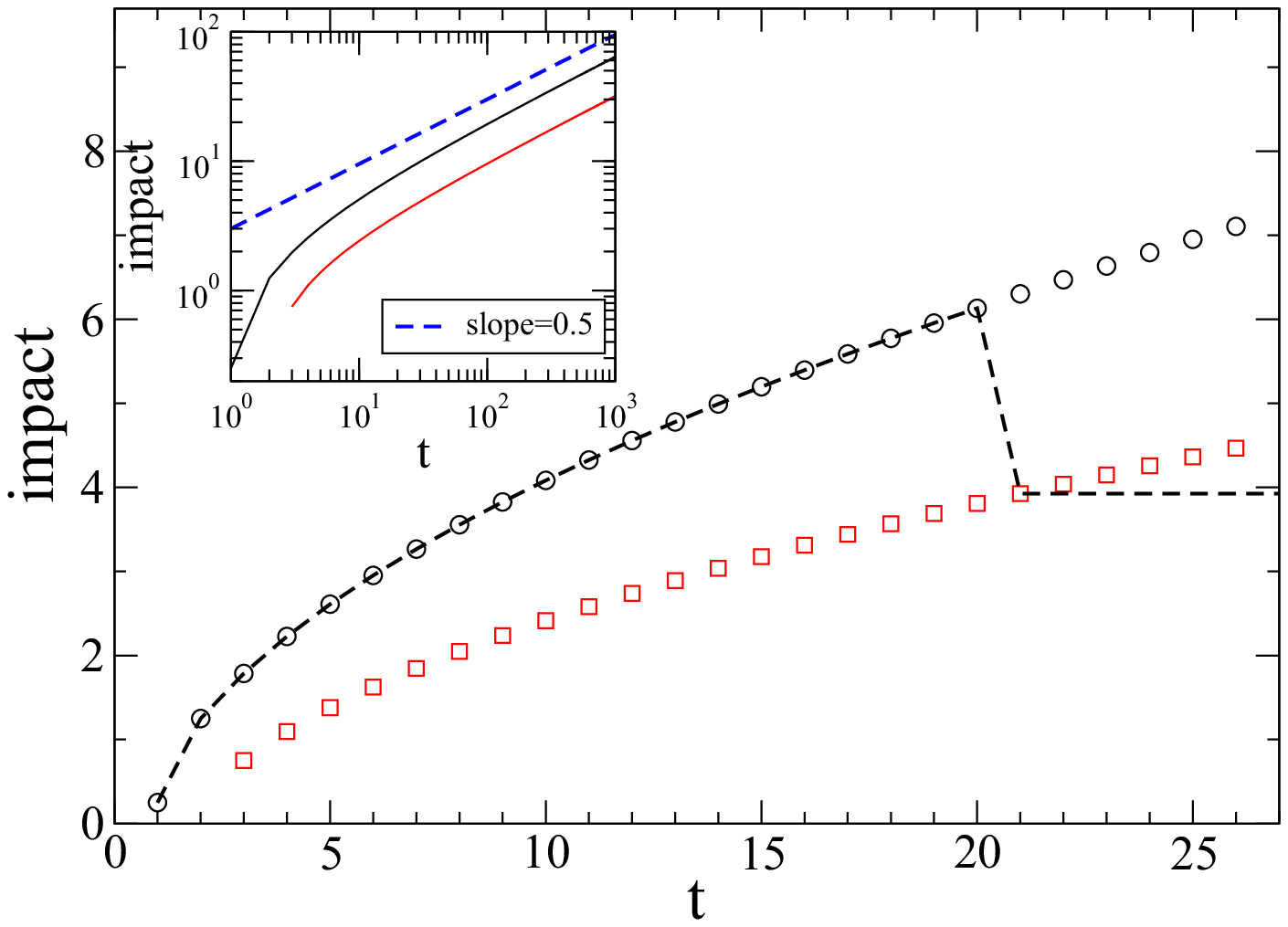}
\includegraphics[scale=0.4]{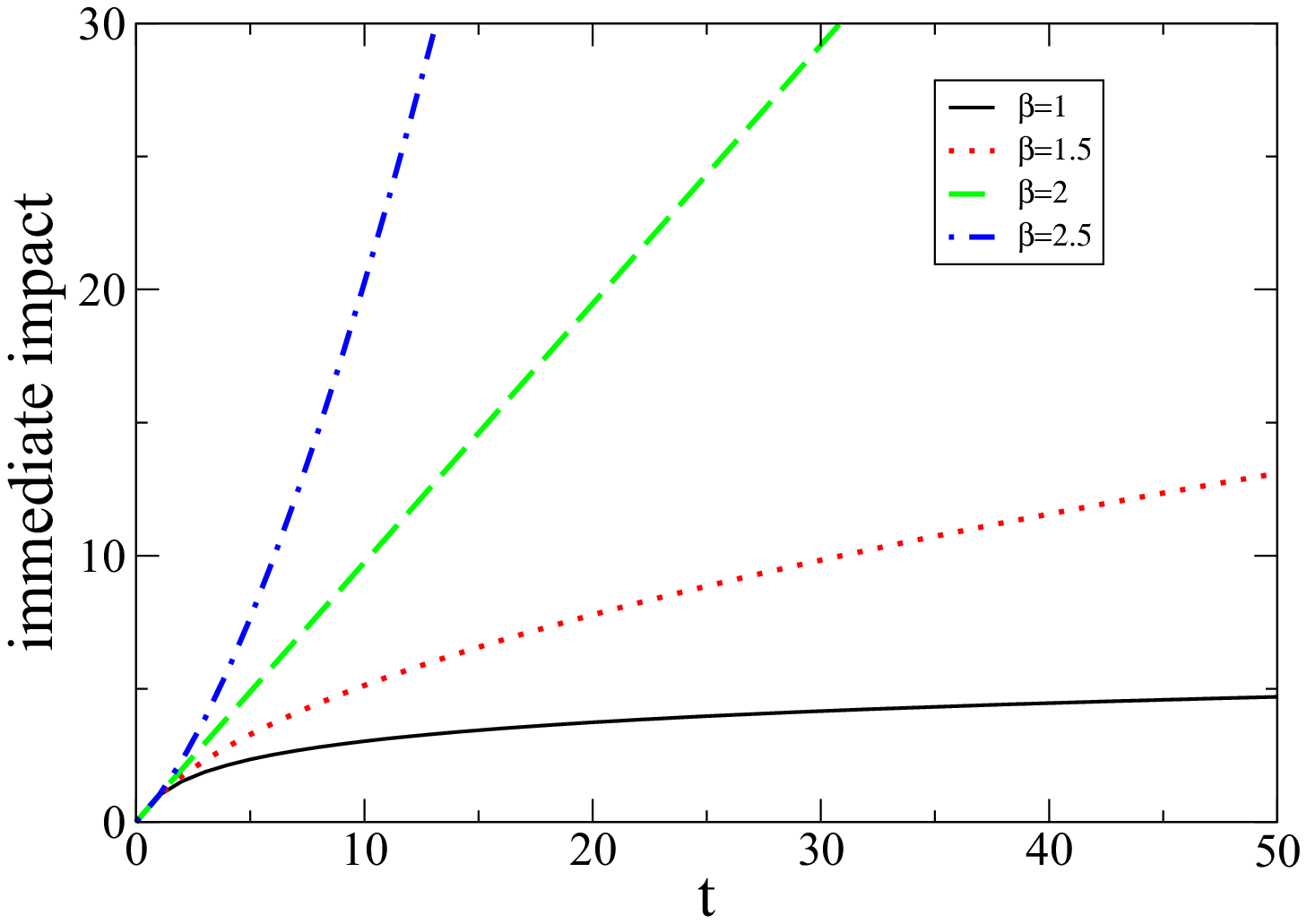}
\end{center}
\caption{An illustration of predicted market impact for Pareto distributed metaorder size. {\it Top panel}: Immediate impact $\mathcal{I}_t$ (black circles) and permanent impact $I_t$ (red squares) for $\beta=1.5$. The dashed line is the price profile of a metaorder of size $N=20$, demonstrating how the price reverts from immediate to permanent impact when metaorder execution is completed.  The inset shows a similar plot in double logarithmic scale for a wider range of sizes (from $N=1$ to $N = 1000$). The blue dashed line is a comparison to the asymptotic square root scaling.  {\it Bottom panel}: Expected immediate impact $\mathcal{I}_t$ as a function of time $t$ for tail exponents $\beta = 1, 1.5, 2$ and $2.5$, illustrating how the impact goes from concave to convex as $\beta$ increases.
 }
\label{numerical}
\end{figure} 
Figure~\ref{numerical} illustrates the reversion process for $\beta=1.5$ and shows how the shape of the impact varies with $\beta$.

\subsubsection{Permanent impact}

The permanent impact under the Pareto assumption is easily computed using Eq.~(\ref{generalPermanentImpact}).
A direct calculation shows that\footnote{
This is the same impact vs. size derived by Gabaix et al. (\citeyear{Gabaix06}).  Their derivation is based on quite different reasoning, and requires mean-variance utility with a linear (rather than the usual quadratic) risk aversion term.  They predict a different permanent impact.}
\begin{equation}
I_{N} \sim \frac{1}{N}\int^N x^{\beta-1}dx =\frac{1}{\beta} N^{\beta-1}.
\label{permanentImpact}
\end{equation}
Eqs.~(\ref{permanentImpact}) and (\ref{temporaryImpact}) imply that the ratio of the permanent to the immediate impact is
\begin{equation}
{I_N \over {\cal I}_N} = \frac{1}{\beta}.
\end{equation}
For example if $\beta=1.5$ the model predicts that on average the permanent impact is equal to two thirds of the maximum immediate impact, i.e., following the completion of a metaorder the price should revert by one third from its peak value.

\subsection{Effect of maximum order size} \label{finiteSize}

In the previous section we have assumed that $N \ll M$, so that we can treat the problem as if $M$ were infinite.  In real markets the maximum order size is probably quite large, a significant fraction of the market capitalization of the asset.  Thus we doubt that the finite support of $N$ has much practical importance, except perhaps for extremely large metaorders.  

From a conceptual point of view, however, having an upper bound on metaorder size creates some interesting effects.  As we have already mentioned, Proposition 1 fails to hold when $M = \infty$, so this must be handled with some care.  In Appendix C we illustrate how the results change when $N \approx M$.  What we observe is that when $N < M/2$ the impact is roughly unchanged from its behavior in the limit $M \to \infty$, but when $N \gtrsim M/2$  the impact becomes highly convex.  This is caused by the fact that the market maker knows the metaorder must end when $t = M$.  Since by definition $\mathcal{P}_M = 0$, the martingale condition requires that $X_{M} = S_M$, i.e. there is no reversion when the metaorder is completed.  This propagates backward and when $N \approx M$ it significantly alters the impact, as seen in Figure~\ref{figimpact}.

Nonetheless, from a practical point of view we do not think this is an important issue, which is why we have relegated the details of the discussion to Appendix C.

\subsection{Relation between volume and information}

One of the main contributions of this paper is to derive a relationship between information and metaorder size.  We now make this more explicit and show that, under the assumption that returns reflect information, it results in the expected relationship between returns and volume. 

In the continuous limit for large $N$ the distributions for information and metaorder size are related by conservation of probability as

\begin{equation}
p(\alpha) =  p_N  \frac{d N}{d \alpha}.
\label{volumeInformation}
\end{equation}
So for example, if the empirical metaorder size is $p_N$ is asymptotically Pareto distributed, as argued in the next section, $p_N \sim N^{-(\beta + 1)}$ and $\alpha = N^{\beta - 1}$ as shown below, we have $p(\alpha) = \alpha^{(1 - 2\beta)/(\beta - 1)}$.  Based on the empirically observed value $\beta \approx 1.5$, this gives $p(\alpha) = \alpha^{-4}$, which means that the cumulative scales as $P(\alpha > x) \sim x^{-3}$.  This is what is typically observed for price returns in American stock markets (Plerou et al., \citeyear{Plerou99}).  This is consistent with (Gabaix, \citeyear{Gabaix03}), with the exception that our analysis is specifically for metaorders.

\section{Discussion}  \label{discussion}

\subsection{Comparison to the Kyle model}

The three types of agents in our model are similar to Kyle's; his informed trader is replaced by our long-term traders, and his noise traders are replaced by our day traders.  In both cases we assume a final liquidation.  There are also several key differences. In our model:
\begin{itemize}
\item
Our long-term traders do not have a monopoly, but rather have common information and compete in setting the size of their orders. Their orders are bundled together and executed as a package by an algorithmic execution service.  These two facts are essential to show that the fair pricing condition is a Nash equilibrium.
\item 
The number of periods $N$ for execution is uncertain, and depends on the information of the long-term traders.  This is important because the martingale condition is based on the market makers' uncertainty about when the metaorder will terminate.
\item
The distribution $p(\alpha)$ is arbitrary (whereas Kyle assumed a normal distribution).  Our key result is that information is almost fully reflected in metaorder size, i.e. that $p_N$ and $p(\alpha)$ are closely related.   This means our results apply to any empirical size distribution $p_N$.
\item
Our day traders respond to ongoing information signals that are permanent in the sense that they affect the final price, in contrast to Kyle's noise traders, who are completely uninformed.   This means that the information in the final liquidation price is incrementally revealed.
\item
The most important difference is that our model has a different purpose.  Kyle assumed an information signal and solved for the optimal strategy to exploit it.  We assume metaorder execution may or may not be going on in the background, and solve for its impact on prices.
\end{itemize}  

\subsection{Empirical implications and tests of the model} \label{empiricalImplications}

The theory presented here makes several predictions with clear empirical implications.  In this section we summarize what these are and outline a few of the problems that are likely to be encountered in empirical testing.  

\begin{enumerate}
\item
The fair pricing condition, Eq.~\ref{fairPricing}, is directly testable, although it requires a somewhat arbitrary choice about when enough time has elapsed since the metaorder has completed for reversion to occur.  (One wants to minimize this time because of the diffusive nature of prices, but one wants to allow enough time to make sure that reversion is complete).
\item
The asymmetric price response predicted by Eq.~\ref{returnRatio} is testable.  However, this only tests the martingale condition, which is the less controversial part of our model.
\item
The equivalence of impact as a function of time and size is directly testable. Under our theory, for $N > t$ the immediate impact from the first $t$ steps is the same, regardless of $N$.   This is in contrast to the Kyle model which predicts linear impact as a function of time, but can explain concavity in size only by postulating variable informativeness of trades vs.  metaorder size.
\item
The prediction of immediate and permanent impact based on $p_N$ is directly testable through Equations~\ref{final} and \ref{generalPermanentImpact}.
\item
If the metaorder distribution is a power law (Pareto distribution), then for large $N$ the immediate impact scales as $\mathcal{I}_t \sim t^{\beta -1}$ and the ratio of the permanent to the immediate impact of the last transaction is $I_N/\mathcal{I}_N = 1/\beta$.  See Section~\ref{paretoSec}.
\end{enumerate}

Prediction (2) has been tested and confirmed by Lillo and Farmer (\citeyear{Lillo03c}), Farmer et al. (\citeyear{Farmer06}) and Gerig (\citeyear{Gerig07}).  
Preliminary results seem to support, or at least not contradict, prediction (5).  The only studies of which we are aware that attempted to fit functional form to the impact of metaorders are by Torre (\citeyear{Torre97}), Almgren et al. \citeyear{Almgren05}, Moro et al. (\citeyear{Moro09}), Toth et al. (\citeyear{Toth11b}), and Bershova and Rakhlin (\citeyear{Bershova13a}). They find immediate impact roughly consistent with a square root functional form.  Moro et al. also tested the ratio of permanent to immediate impact and found $0.51$ for the Spanish stock market and $0.69$ for the London stock market, with large error bars.  Very recently (subsequent to the appearance of our paper in preprint form), Bershova and Rakhlin (\citeyear{Bershova13a}) and Gomes and Waelbroeck (\citeyear{Gomes13}) each investigated separate proprietary sets of large metaorders on US equity markets. They independently verify the fair pricing condition.  Moreover Bershova and Rakhlin (\citeyear{Bershova13a}) show that the permanent impact is approximated by a square root function of execution time and that its ratio to the immediate impact is close to $2/3$. These last studies are the strongest piece of direct evidence for our theory.

\subsection{Information revelation}

Though the market makers in our model are uncertain whether or not a metaorder is present, if it is present, they know when its execution begins and ends.  The ability to detect metaorders from imbalances in order flow using brokerage codes has been demonstrated (see (Vaglica et al. \citeyear{Vaglica08}),  (Toth et al \citeyear{Toth10})).   A recent study of metaorders based on brokerage code information found average participation rates of $17\%$ for the Spanish stock market (BME) and $34\%$ for the London Stock Market\footnote{
Participation rate is defined as the fraction of trades that a given agent participates in.},
for metaorders whose average size was just under $100$ in both markets, making such metaorders difficult to hide.  The detection problem introduces uncertainties in starting and stopping times that may affect shape of the price impact.  

%

\subsection{Final thoughts}

The traditional view in finance is that market impact is just a reflection of information.  This point of view often goes a step further and postulates that the functional form of impact is determined by behavioral and institutional factors, such as how informed the agents are who trade with a given volume.  This hypothesis is difficult to test because it is inherently complicated and information is difficult to measure independently of impact.   Within the framework developed here, such anomalies would violate the fair pricing condition.

In this paper we embrace the view that impact reflects information, but we show how at equilibrium the trading volume reflects the underlying information and makes it possible to compute the impact.  The metaorder size distribution determines the shape of the impact but does not set its scale.   Metaorder size has the important advantage of being a measurable quantity, and thus predictions based on it are much more testable than those based directly on information.  

The fair pricing condition that we have derived here may well hold on its own, even without informational efficiency.  This could be true for purely behavioral reasons:  The fair pricing condition holds because it can be measured reliably, and both parties view it as fair.  Thus while the main results here are consistent with rationality, they do not necessarily depend on it.

We provide an example solution for the Pareto distribution for metaorder size because we believe that the evidence supports this hypothesis.  This gives the simple result that the impact is a power law of the form $\mathcal{I}_t \sim t^{\beta - 1}$, and the ratio of permanent impact to the temporary impact of the last transaction is $I_N/\mathcal{I}_N = 1/\beta$.   However, the  bulk of our results do not depend on this assumption.  Thus the reader who is skeptical about power laws may simply view the results for the Pareto distribution as a worked example.

The strength of our approach is its empirical predictions.  Because these involve explicit functional relationships between observable variables they are strongly falsifiable in the Popperian sense.  A preliminary empirical analysis seems to support the theory, but the statistical analysis so far remains inconclusive. We look forward to more rigorous empirical tests.  

\section*{Acknowledgments} We would like to acknowledge conversations with Jean-Philippe Bouchaud and a very helpful discussion by Ionid Rosu.  This work was supported by PRIN project 2007TKLTSR  ``Computational markets design and agent-based models of trading behavior" and National Science Foundation grants 0624351 and 0965673.  FL acknowledges financial support from the grant "Progetto Interno Lillo 2010" by Scuola Normale Superiore di Pisa.
\bibliographystyle{jedc}

\begin{thebibliography}{}

\end{thebibliography}


\begin{thebibliography}{}

\bibitem[\protect\citeauthoryear{Almgren}{2003}]{Almgren03}
Almgren, R.F., 2003.
\newblock Optimal execution with nonlinear impact functions and
  trading-enhanced risk.
\newblock Applied Mathematical Finance, 10{,} 1--18.

\bibitem[\protect\citeauthoryear{Almgren and Chriss}{1999}]{Almgren99}
Almgren, R.F. and Chriss, N., 1999.
\newblock Value under liquidation.
\newblock Risk, 12{,} 61--63.

\bibitem[\protect\citeauthoryear{Almgren and Chriss}{2000}]{Almgren00}
Almgren, R.F. and Chriss, N., 2000.
\newblock Optimal execution of portfolio transactions.
\newblock Journal of Risk, 3{,} 5--39.

\bibitem[\protect\citeauthoryear{Almgren \bgroup et al.\egroup
  }{2005}]{Almgren05}
Almgren, R.F., Thum, C. and Hauptmann, H.~L., 2005.
\newblock Direct estimation of equity market impact.
\newblock Risk.

\bibitem[\protect\citeauthoryear{Back and Baruch}{2007}]{Back07}
Back, Kerry and Baruch, Shmuel, 2007.
\newblock Working orders in limit order markets and floor exchanges.
\newblock Journal of Finance, 62{,} 1589--1621.

\bibitem[\protect\citeauthoryear{Beran}{1994}]{Beran94}
Beran, J., 1994.
\newblock Statistics for Long-Memory Processes.
\newblock New York: Chapman \& Hall.

\bibitem[\protect\citeauthoryear{Berk and Green}{2004}]{Berk04}
Berk, J.B. and Green, R.C, 2004.
\newblock Mutual fund flows and performance in rational markets.
\newblock Journal of Political Economy, 112{,} 1269--1295.

\bibitem[\protect\citeauthoryear{Bernhardt \bgroup et al.\egroup
  }{2002}]{Bernhardt02}
Bernhardt, Dan, Hughson, Eric and Naganathan, Girish, 2002.
\newblock Cream-skimming and payment for order flow.
\newblock Technical Report. University of Illinois.


\bibitem[\protect\citeauthoryear{Bershova and Rakhlin}{2013a}]{Bershova13a}
Bershova, N. and Rakhlin, D., 2013
\newblock The Non-Linear Market Impact of Large Trades: Evidence from Buy-Side Order Flow
\newblock http://ssrn.com/abstract=2197534.
  
  \bibitem[\protect\citeauthoryear{Bershova and Rakhlin}{2013b}]{Bershova13b}
Bershova, N. and Rakhlin, D., 2013
\newblock High-Frequency Trading and Long-Term Investors: A View from the Buy-Side
\newblock http://ssrn.com/abstract=2066884.

\bibitem[\protect\citeauthoryear{Bertismas and Lo}{1998}]{Bertismas98}
Bertismas, D. and Lo, A., 1998.
\newblock Optimal control of execution costs.
\newblock Journal of Financial Markets, 1{,} 1--50.

\bibitem[\protect\citeauthoryear{Bouchaud \bgroup et al.\egroup
  }{2009}]{Bouchaud08b}
Bouchaud, J-P, Farmer, J.~Doyne and Lillo, F., 2009.
\newblock How markets slowly digest changes in supply and demand, in:
\newblock Hens, T. and Schenk-Hoppe, K. (Eds.), Handbook of Financial Markets:
  Dynamics and Evolution.
\newblock Elsevier, {pp.} 57--160.

\bibitem[\protect\citeauthoryear{Bouchaud \bgroup et al.\egroup
  }{2004}]{Bouchaud04}
Bouchaud, J-P., Gefen, Y., Potters, M. and Wyart, M., 2004.
\newblock Fluctuations and response in financial markets: The subtle nature of
  ``random" price changes.
\newblock Quantitative Finance, 4{,} 176--190.

\bibitem[\protect\citeauthoryear{Bouchaud \bgroup et al.\egroup
  }{2006}]{Bouchaud04b}
Bouchaud, J-P., Kockelkoren, J. and Potters, M., 2006.
\newblock Random walks, liquidity molasses and critical response in financial
  markets.
\newblock Quantitative Finance, 6{,} 115--123.

\bibitem[\protect\citeauthoryear{Challet}{2007}]{Challet07b}
Challet D., 2007.
\newblock The demise of constant price impact functions and single-time step
  models of speculation.
\newblock Physica A., 382{,} 29--35.

\bibitem[\protect\citeauthoryear{Chan and Lakonishok}{1993}]{Chan93}
Chan, L.~K.C. and Lakonishok, J., 1993.
\newblock Institutional trades and intraday stock price behavior.
\newblock Journal of Financial Economics, 33{,} 173--199.

\bibitem[\protect\citeauthoryear{Chan and Lakonishok}{1995}]{Chan95}
Chan, L.~K.C. and Lakonishok, J., 1995.
\newblock The behavior of stock prices around institutional trades.
\newblock The Journal of Finance, 50{,} 1147--1174.

\bibitem[\protect\citeauthoryear{Chordia and Subrahmanyam}{2004}]{Chordia04}
Chordia, T. and Subrahmanyam, A., 2004.
\newblock Order imbalance and individual stock returns: Theory and evidence.
\newblock Journal of Financial Markets, 72{,} 485--518.

\bibitem[\protect\citeauthoryear{Ding \bgroup et al.\egroup }{1993}]{Ding93}
Ding, Z., Granger, C. W.~J. and Engle, R.~F., 1993.
\newblock A long memory property of stock returns and a new model.
\newblock Journal of Empirical Finance, 1{,} 83--106.

\bibitem[\protect\citeauthoryear{Donier}{2012}]{Donier12}
Donier, J., 2012
\newblock Market Impact with Autocorrelated Order Flow under Perfect Competition.
\newblock http://arxiv.org/abs/1212.4770

\bibitem[\protect\citeauthoryear{Eisler and Kertecz}{2006}]{Eisler06}
Eisler, Z. and Kertesz, J., 2006.
\newblock Size matters, some stylized facts of the market revisited.
\newblock European Journal of Physics B, 51{,} 145--154.

\bibitem[\protect\citeauthoryear{Engle \bgroup et al.\egroup }{2008}]{Engle08}
Engle, R., Ferstenberg, R. and Russel, J., 2008.
\newblock Measuring and modeling execution cost and risk.
\newblock Technical Report 08-09. University of Chicago.

\bibitem[\protect\citeauthoryear{Evans and Lyons}{2002}]{Evans02}
Evans, M. D.~D. and Lyons, R.~K., 2002.
\newblock Order flow and exchange rate dynamics.
\newblock Journal of Political Economy, 110{,} 170--180.

\bibitem[\protect\citeauthoryear{Farmer \bgroup et al.\egroup
  }{2004}]{Farmer04b}
Farmer, J.~D., Gillemot, L., Lillo, F., Mike, S. and Sen, A., 2004.
\newblock What really causes large price changes?
\newblock Quantitative Finance, 4{,} 383--397.

\bibitem[\protect\citeauthoryear{Farmer \bgroup et al.\egroup
  }{2005}]{Farmer05}
Farmer, J.~D., Patelli, P. and Zovko, Ilija, 2005.
\newblock The predictive power of zero intelligence in financial markets.
\newblock Proceedings of the National Academy of Sciences of the United States
  of America, 102{,} 2254--2259.

\bibitem[\protect\citeauthoryear{Farmer \bgroup et al.\egroup
  }{2006}]{Farmer06}
Farmer, J.D., Gerig, A., Lillo, F. and Mike, S., 2006.
\newblock Market efficiency and the long-memory of supply and demand: Is price
  impact variable and permanent or fixed and temporary?
\newblock Quantitative Finance, 6{,} 107--112.

\bibitem[\protect\citeauthoryear{Gabaix \bgroup et al.\egroup
  }{2003}]{Gabaix03}
Gabaix, X., Gopikrishnan, P., Plerou, V. and Stanley, H.~E., 2003.
\newblock A theory of power-law distributions in financial market fluctuations.
\newblock Nature, 423{,} 267--270.

\bibitem[\protect\citeauthoryear{Gabaix \bgroup et al.\egroup
  }{2006}]{Gabaix06}
Gabaix, X., Gopikrishnan, P., Plerou, V. and Stanley, H.E., 2006.
\newblock Institutional investors and stock market volatility.
\newblock Quarterly Journal of Economics, 121{,} 461--504.

\bibitem[\protect\citeauthoryear{Gatheral}{2010}]{Gatheral10}
Gatheral, J., 2010.
\newblock No-dynamic-arbitrage and market impact.
\newblock Quantitative Finance, 10{,} 749--759.

\bibitem[\protect\citeauthoryear{Gerig}{2007}]{Gerig07}
Gerig, A., 2007.
\newblock A theory for market impact: How order flow affects stock price.
\newblock {Ph.D.} Thesis. University of Illinois.

\bibitem[\protect\citeauthoryear{Gerig}{2011}]{Gerig11}
Gerig, A., Farmer, J.~D., and Lillo, F., 2011.
\newblock How Prices Respond to Worked Orders.
\newblock Working Paper.

\bibitem[\protect\citeauthoryear{Gillemot}{2006}]{Gillemot06}
Gillemot, L., Farmer, J.~D., and Lillo, F., 2006.
\newblock There’s more to volatility than volume.
\newblock Quantitative Finance, 6{,} 371--384.

\bibitem[\protect\citeauthoryear{Glosten}{2003}]{Glosten03}
Glosten, L.~R., 2003.
\newblock Discriminatory limit order books, uniform price clearing and
  optimality.
\newblock Technical Report. Columbia.

\bibitem[\protect\citeauthoryear{Glosten}{1994}]{Glosten94}
Glosten, L.~R., 1994.
\newblock Is the electronic limit order book inevitable?
\newblock Journal of Finance, 49{,} 1127--1161.

\bibitem[\protect\citeauthoryear{Gomes and Waelbroeck}{2013}]{Gomes13}
Gomes, C. and Waelbroeck, H., 2013.
\newblock Is Market Impact a Measure of the Information Value of Trades? Market Response to Liquidity vs Informed Trades.
\newblock http://ssrn.com/abstract=2291720.

\bibitem[\protect\citeauthoryear{Gopikrishnan \bgroup et al.\egroup
  }{2000}]{Gopikrishnan00}
Gopikrishnan, P., Plerou, V., Gabaix, X. and Stanley, H.~E., 2000.
\newblock Statistical properties of share volume traded in financial markets.
\newblock Physical Review E, 62{,} R4493--R4496; Part A.

\bibitem[\protect\citeauthoryear{Hasbrouck}{1991}]{Hasbrouck91}
Hasbrouck, J., 1991.
\newblock Measuring the information content of stock trades.
\newblock The Journal of Finance, 46{,} 179--207.

\bibitem[\protect\citeauthoryear{Hausman \bgroup et al.\egroup
  }{1992}]{Hausman92}
Hausman, J.~A., Lo, A.~W. and Mackinlay, A.~C., 1992.
\newblock An ordered probit analysis of transaction stock prices.
\newblock Journal of Financial Economics, 31{,} 319--379.

\bibitem[\protect\citeauthoryear{Hopman}{2006}]{Hopman02}
Hopman, C., 2006.
\newblock Do supply and demand drive stock prices?
\newblock Quantitative Finance; To appear.

\bibitem[\protect\citeauthoryear{Huberman and Stanzl}{2004}]{Huberman04b}
Huberman, G. and Stanzl, W., 2004.
\newblock Price manipulation and quasi-arbitrage.
\newblock Econometrica, 72{,} 1247--1275.

\bibitem[\protect\citeauthoryear{Keim and Madhavan}{1996}]{Keim96}
Keim, D.~B. and Madhavan, A., 1996.
\newblock The upstairs market for large-block transactions:analysis and
  measurement of price effects.
\newblock The Review of Financial Studies, 9{,} 1--36.

\bibitem[\protect\citeauthoryear{Kempf and Korn}{1999}]{Kempf99}
Kempf, A. and Korn, O., 1999.
\newblock Market depth and order size.
\newblock Journal of Financial Markets, 2{,} 29--48.

\bibitem[\protect\citeauthoryear{Kyle}{1985}]{Kyle85}
Kyle, A.~S., 1985.
\newblock Continuous auctions and insider trading.
\newblock Econometrica, 53{,} 1315--1335.

\bibitem[\protect\citeauthoryear{Lillo and Farmer}{2004}]{Lillo03c}
Lillo, F. and Farmer, J.~D., 2004.
\newblock The long memory of the efficient market.
\newblock Studies in Nonlinear Dynamics \& Econometrics, 8{,}1.

\bibitem[\protect\citeauthoryear{Lillo \bgroup et al.\egroup }{2003}]{Lillo03d}
Lillo, F., Farmer, J.~D. and Mantegna, R.~N., 2003.
\newblock Master curve for price impact function.
\newblock Nature, 421{,} 129--130.

\bibitem[\protect\citeauthoryear{Lillo \bgroup et al.\egroup }{2005}]{Lillo05b}
Lillo, F., Mike, S. and Farmer, J.~D., 2005.
\newblock Theory for long memory in supply and demand.
\newblock Physical Review E, 7106{,} 066122.

\bibitem[\protect\citeauthoryear{Moro \bgroup et al.\egroup }{2009}]{Moro09}
Moro, E., Moyano, L.~G., Vicente, J., Gerig, A., Farmer, J.~D., Vaglica, G.,
  Lillo, F. and Mantegna, R.N., 2009.
\newblock Market impact and trading profile of hidden orders in stock markets.
\newblock Physical Review E., 80{,} 066102.

\bibitem[\protect\citeauthoryear{Obizhaeva and Wang}{2005}]{Obizhaeva05}
Obizhaeva, A.A. and Wang, J., 2005.
\newblock Optimal trading strategy and supply/demand dynamcis.
\newblock Technical Report. AFA 2006 Boston Meetings Paper.

\bibitem[\protect\citeauthoryear{Plerou \bgroup et al.\egroup
  }{1999}]{Plerou99}
Plerou, V., Gopikrishnan, P., Amaral, L. A.~N., Meyer, M. and Stanley, H.~E.,
  1999.
\newblock Scaling of the distribution of price fluctuations of individual
  companies.
\newblock Physical Review E, 60{,} 6519--6529; Part A.

\bibitem[\protect\citeauthoryear{Plerou \bgroup et al.\egroup
  }{2002}]{Plerou02}
Plerou, V., Gopikrishnan, P., Gabaix, X. and Stanley, H.~E., 2002.
\newblock Quantifying stock price response to demand fluctuations.
\newblock Physical Review E, 66{,} article no. 027104.

\bibitem[\protect\citeauthoryear{Potters and Bouchaud}{2003}]{Potters03}
Potters, M. and Bouchaud, J-P., 2003.
\newblock More statistical properties of order books and price impact.
\newblock Physica A, 324{,} 133--140.

\bibitem[\protect\citeauthoryear{Racz \bgroup et al.\egroup }{2009}]{Racz07}
Racz, E., Eisler, Z. and Kertesz, J., 2009.
\newblock Comment on `tests of scaling and universality.' by plerou and
  stanley.
\newblock Technical Report.

\bibitem[\protect\citeauthoryear{Schwartzkopf and
  Farmer}{2010}]{Schwartzkopf10}
Schwartzkopf, Y. and Farmer, J.D., 2010.
\newblock Technical Report.
\newblock For a preliminary report, see Y. Schwartzkopf's Caltech Ph.D thesis,
  Complex Phenomena in Social and Financial Systems: From bird population
  growth to the dynamics of the mutual fund industry.

\bibitem[\protect\citeauthoryear{Torre}{1997}]{Torre97}
Torre, N., 1997.
\newblock BARRA Market Impact Model Handbook.
\newblock Berkeley: BARRA Inc.

\bibitem[\protect\citeauthoryear{Toth \bgroup et al.\egroup }{2011a}]{Toth11b}
Toth, B., Lemperiere, Y., Deremble, C., de~Lataillade, J., Kockelkoren, J. and
  Bouchaud, J-P., 2011.
\newblock Anomalous price impact and the critical nature of liquidity in
  financial markets.
\newblock Physical Review X, 1{,} 021006.

\bibitem[\protect\citeauthoryear{Toth \bgroup et al.\egroup }{2010}]{Toth10}
Toth, B., Lillo, F. and Farmer, J.D., 2010.
\newblock Segmentation algorithm for non-stationary compound Poisson processes.
With an application to inventory time series of market members in a financial market
\newblock European Physical Journal B, 78{,} 235--243.

\bibitem[\protect\citeauthoryear{Toth \bgroup et al.\egroup }{2011b}]{Toth11c}
Toth, B., Palit, I., Lillo, F. and Farmer, J.~D., 2011.
\newblock Why is order flow so persistent?
\newblock arXiv:1108.1632.

\bibitem[\protect\citeauthoryear{Vaglica}{2008}]{Vaglica08}
Vaglica, G., 2008.
\newblock Scaling laws of strategic behavior and specialization of strategies
  in agent dynamics of a financial market.
\newblock {Ph.D.} Thesis. University of Palermo.

\bibitem[\protect\citeauthoryear{Vaglica \bgroup et al.\egroup
  }{2008}]{Vaglica07}
Vaglica, G., Lillo, F., Moro, E. and Mantegna, R., 2008.
\newblock Scaling laws of strategic behavior and size heterogeneity in agent
  dynamics.
\newblock Physical Review E., 77{,} 036110.

\bibitem[\protect\citeauthoryear{Viswanathan and Wang}{2002}]{Viswanathan02}
Viswanathan, J. and Wang, James J.~D., 2002.
\newblock Market architecture: Limit order books versus dealership markets.
\newblock Journal of Financial Markets, 5{,} 127--167.

\bibitem[\protect\citeauthoryear{Weber and Rosenow}{2006}]{Weber04}
Weber, P. and Rosenow, B., 2006.
\newblock Large stock price changes: volume or liquidity?
\newblock Quantitative Finance, 6{,} 7--14.

\bibitem[\protect\citeauthoryear{Wyart \bgroup et al.\egroup }{2006}]{Wyart06}
Wyart, M., Bouchaud, J.-P., Kockelkoren, J., Potters, M. and Vettorazzo, M.,
  2006.
\newblock Relation between bid-ask spread, impact and volatility in double
  auction markets.
\newblock Technical Report.

\bibitem[\protect\citeauthoryear{Zhang}{1999}]{Zhang99}
Zhang, Y.~C., 1999.
\newblock Toward a theory of marginally efficient markets.
\newblock Physica A, 269{,} 30--44.

\end{thebibliography}

\newpage

\appendix
\section{Proofs of the propositions}
 
 {\bf Proposition 1.} {\it The martingale condition implies zero overall profits, i.e.} 
\begin{equation}
E_1[N\pi_N]\equiv\sum_{N=1}^M p_N N \pi_N=0.
\label{appeq1}
\end{equation}

{\bf Proof.} Given the definition of $R^+_t$ and $R^-_t$ we can write the prices as
\begin{eqnarray}
S_t=X_0+\sum_{i=0}^{t-1} R^+_i \\
X_{t}=X_0+\sum_{i=0}^{t-1} R^+_i-R^-_t
\end{eqnarray}  
With these expressions for $N<M$, $\pi_N$ can be rewritten as
\begin{eqnarray}
\pi_N= \frac{1}{N} \sum_{t=1}^N S_t - X_{N}=R^-_N-\frac{1}{N}\sum_{i=1}^{N-1}iR^+_i=\nonumber \\
=\frac{1}{p_N}R^+_N \sum_{i=N+1}^M p_i-\frac{1}{N}\sum_{i=1}^{N-1}iR^+_i
\label{profpershare}
\end{eqnarray}
 where in the last equality we have used the martingale condition of Eq. (\ref{returnRatio}).
 For $N=M$ the profit per share is
 \begin{eqnarray}
 \pi_M= \frac{1}{M} \sum_{i=1}^M S_i - X_{M}=\frac{1}{M} \sum_{i=1}^M S_i - S_{M}=\nonumber \\
 =-\frac{1}{M}\sum_{i=1}^{M-1}iR^+_i
 \end{eqnarray}
 By substituting these two last expressions in Eq. (\ref{appeq1}) we obtain
\begin{eqnarray}
E_1[N\pi_N]=\sum_{N=1}^{M-1}NR^+_N \sum_{i=N+1}^{M} p_i-\sum_{N=1}^{M-1}p_N\sum_{i=1}^{N-1}iR^+_i-p_M\sum_{i=1}^{M-1} iR^+_i =\nonumber \\
=\sum_{N=1}^{M-1}NR^+_N \sum_{j=N+1}^{M} p_j-\sum_{N=1}^{M}p_N\sum_{i=1}^{N-1}iR^+_i
\end{eqnarray}
By explicitly computing the coefficients of each $R^+_i$, it is easy to show they vanish for each $i$, i.e. $E_1[N\pi_N]=0$.

{\bf Infinite support}.  This proposition does not hold when $\alpha$ has infinite support. in order to show this let us consider the expected profit for orders of length between $N=1$ and $N=\bar N$. The following proposition holds:

 {\bf Proposition 1$'$.} {\it The martingale condition for all intervals implies that for any integer $\bar{N} \ge 1$,
\begin{equation}
\sum_{N=1}^{\bar N} p_N N\pi_N =\left(\sum_{i=\bar N+1}^M p_i\right) \left(\sum_{i=1}^{\bar N} iR^+_i\right) \ge 0.
\label{partsum2}
\end{equation}
This equation holds both for finite and infinite support (i.e. $M$ can be finite or infinite).}

{\bf Proof.} 
From the equation (\ref{profpershare}) in the previous proposition, we know that martingale condition allows us to write the profit per lot traded as
\begin{equation}
\pi_N=\frac{1}{p_N}R^+_N \sum_{i=N+1}^M p_i-\frac{1}{N}\sum_{i=1}^{N-1}iR^+_i\nonumber
\end{equation}
Therefore the expected profit for orders shorter or equal to $\bar N<M$ is 
\begin{eqnarray}
\sum_{N=1}^{\bar N} p_N N\pi_N =\sum_{N=1}^{\bar N}N R^+_N\sum_{i=N+1}^{M}p_i-\sum_{N=1}^{\bar N}p_N\sum_{i=1}^{N-1}iR^+_i=\nonumber \\
\left(\sum_{i=\bar N+1}^M p_i\right) \left(\sum_{i=1}^{\bar N} iR^+_i\right) 
\end{eqnarray}
This is equal to the quantity in Eq. (\ref{partsum2}). Moreover it is clear that both terms in brackets are non-negative.  This means that the market maker typically makes profits on short metaorders. 

If the support of $p_N$ is infinite then the martingale condition at all intervals implies that 
\begin{equation}
E_1[N\pi_N]\equiv\sum_{N=1}^\infty p_N N\pi_N=\lim_{\bar N\to\infty}  \left(\sum_{i=\bar N+1}^\infty p_i\right) \left(\sum_{i=1}^{\bar N} iR^+_i\right).
\end{equation}
In the infinite support case the behavior of the limit in the last term of the above expression depends on the asymptotic behavior of $p_N$ and $R^+_N$ for large $N$. This is due to the fact that for large $\bar N$ the first term in brackets goes to zero while the second term diverges.   It is possible to construct examples where $E_1[N\pi_N]$ goes to zero, to a finite value, or diverges. This result shows that in the infinite support case the martingale condition does not imply zero overall immediate profits.

\medskip

{\bf Proposition 2.}  {\it If the second derivative of the immediate impact $\mathcal{I}_t$ is bounded strictly below zero, in the limit where the number of informed traders $\mathcal{K} \to \infty$, any Nash equilibrium must satisfy the fair pricing condition $\pi_N = 0$ for $1 < N < M$.  On average market makers profit from orders of length one and take (equal and opposite) losses from orders of length $M$.}

The strategy of the proof is to show that if $\pi_N \ne 0$ for some $N$, the long-term traders would have an incentive to change strategy, so the equilibrium must satisfy $\pi_N = 0$.

A long-term trader $k$ receives the signal $\alpha$ and buys $n_k$ shares at an average price $\sum_{t=1}^{N} \tilde{S}_t/N$.  After averaging over $\eta_t$ the expected average transaction price is $\sum_{t=1}^{N} S_t/N$ and the expected final price is $X_{N} = X_0 + \alpha$.  The expected profit is therefore,
\[
\Pi_k(\alpha) = n_k \left(X_0 + \alpha - {1 \over N}  \sum_{t=1}^{N}
S_t \right) = -n_k \pi(\alpha) \].

If the long-term trader $k$ increases her order size by one lot while all others hold their order size constant, then her expected profit becomes
$$ \Pi'_{k}(\alpha) = (n_k+1)  \left(X_0 + \alpha - {1 
\over {N+1}}  \sum_{t=1}^{N+1}  S_t \right),$$
which after a little algebra can be rewritten,
$$ \Pi'_{k}(\alpha) = -(n_k+1)\pi(\alpha) - \left({{n_k+1} \over {N+1}}\right)   \left(S_{N+1} - {1 
\over {N}}  \sum_{t=1}^{N}  S_t \right).$$
The expected change in profit is therefore, 
\begin{equation}
\Delta \Pi = \Pi'_{k}(\alpha)- \Pi_{k}(\alpha) =  -\pi(\alpha) - \left({{n_k+1} \over {N+1}}\right)
\left(S_{N+1}-{1 \over N} \sum_{t=1}^{N}  S_t \right).
\label{deltaProfit}
\end{equation}
The first term on the right ($-\pi(\alpha)$) represents the additional profit to the long-term trader if it were possible to trade one extra lot at the same average price, and the second term represents the reduction in profit because the average price increases due to the extra lot.  

In the limit as $\mathcal{K}$ is large the long-term trader's own order is a small fraction of the aggregated trade, $n_k \ll N$.  The second term vanishes in the large $N$ limit if
\[
\lim_{N \to \infty} \left({{n_k+1} \over {N+1}}\right)
\left(S_{N+1}-{1 \over N} \sum_{t=1}^{N}  S_t \right) = 0.
\]
This is true providing the second derivative of the function $S_t$ is bounded strictly below zero\footnote{
When the distribution of metaorders is a power law with Pareto exponent $\beta$, Proposition 2 leads to a power law impact function with exponent $\beta-1$, so our assumption on the second derivative of immediate impact leads to a self-consistent theory as long as $\beta<2$. For metaorder distributions with thinner tails than this, the fair pricing condition could be imposed as a hypothesis but it would not follow from a Nash equilibrium. Empirical observations cluster near $\beta=1.5$.}.
Thus in this limit $\Delta \Pi  = -\pi(\alpha) > 0$ and the candidate equilibrium fails because the long-term traders have an incentive to deviate. Similarly if $\pi(\alpha) > 0$ the long-term traders take a loss which can be reduced by trading less.

When $\pi(\alpha)=0$ (and as before $1 < N < M$) no long-term trader has an incentive to change her order size.   This is clear since in Eq.~(\ref{deltaProfit}) with $\pi(\alpha) = 0$ the change in profit is given by the second term alone, which is always negative.  A similar calculation shows that this is also true for decreasing order size, i.e. when $\pi(\alpha) = 0$, $n_k \to n_k - 1$ causes $\Delta \Pi < 0$ . 

To show that this implies $\pi_N=0$ for all $N$, we first discretize $\alpha$ into bins labeled by the mode of $p(N | \alpha)$: $\alpha \in \alpha_i \iff Mode(p(N | \alpha))=i$. Then, we invert the set of homogeneous linear equations $\pi(\alpha_i) = \sum_{N=1}^{M} p(N | \alpha_i) \pi_N = 0$ for $1<i<M$. Since the mode of $p(N | \alpha_i)$ is equal to $i$, the matrix $M_{iN}=p(N | \alpha_i)$ is diagonally-dominant, therefore by the Levy-Desplanques theorem it is also non-singluar. It follows that $\pi_N=0$ for all $N$.

The cases $N = 1$ and $N = M$ have to be examined separately because our equilibrium argument relied on the ability to increase or reduce $n_k$, which is not possible at the boundaries. Long-term traders will rationally abstain from taking a loss on metaorders of length $N = 1$ by simply not participating when they receive $\alpha$ signals that are too weak; thus, the trading volume at $N = 1$ is due entirely to the day trader.  Similarly, although $\pi_M < 0$, the long-term traders are unable to improve their profits by trading more, since we have bounded the total amount an individual can trade at $M/\mathcal{K}$ so they are blocked from further increase.  

The fair pricing condition for $N = 1$ and $N = M$ would also be incompatible with the martingale condition and informational efficiency (i.e. with the conditions on the final price).  For $N = 1$ the market makers' profit is
 \[ 
 \pi_1 = S_1 - X_1 = R^-_1,
 \]
 and from (Eq.~\ref{shortterm}) the martingale condition is
 $${\cal P}_1 (S_2 - S_1) + (1 - {\cal P}_1)(X_1 - S_1) = {\cal P}_1 (S_2 - X_1) + X_1 - S_1 = 0.$$
Thus if $\mathcal{P}_1 \ne 0$, satisfaction of both the martingale condition and the fair pricing condition\footnote{
In different terms, the incompatibility of fair pricing and the martingale condition was pointed out by Glosten (\citeyear{Glosten94}).}
requires that $S_1 = S_2 = X_1$, or equivalently that $R^-_1 = R^+_1$ = 0.  In other words, if both conditions are satisfied then both the permanent and the temporary impact on the first step are identically zero, which would violate informational efficiency since $\alpha > 0$.  In Section~\ref{impactSolution} we show by construction that this holds for all $N$, i.e. it is clear in Eq.~(\ref{final}) and (\ref{generalPermanentImpact}) that the impacts $\mathcal{I}_t$ and $I_N$ are identically zero if $R^-_1 = 0$.  To have sensible impact functions we must have $R^-_1 = \pi_1 > 0$, which means that market making is profitable on the first timestep\footnote{
One might be tempted to naively conclude that market makers can defect from the equilibrium by simply trading orders only of length $1$, so that they always make a profit.  This is false:  The profit from a metaorder of length one is $\pi_1 = S_1 - (X_0 + \alpha)$, where $\alpha$ is a small number.  In contrast, if a market maker participates only in the first trade of a large metaorder, her profit is $\pi'_1 = S_1 - (X_0 + \alpha')$, where $\alpha'$ is a large number.  Thus while $\pi_1 > 0$, $\pi'_1 < 0$.}.

Similarly if $N = M$ the martingale condition implies $S_M = X_{M}$, i.e. no reversion, and since $S_t$ is an increasing function the market maker takes a loss
\[
M \pi_M = \sum_{t=1}^M S_t - M X_{M} = \sum_{i=1}^{M} (S_i - S_M) < 0.
\]
Assuming $\pi_N = 0$ for $1 < N < M$, the market makers' profit $\pi_1$ and loss $M \pi_M$ are related by Eq.~(\ref{breakevenonaverage}) as
\begin{equation}
\pi_1 p_1+M \pi_M p_M=0.
\end{equation}
For realistic size distributions we expect metaorders of size one to be much more common than those of size $M$, i.e. $p_1 \gg p_M$.  The ratio of the total profits is
\[
-\frac{M \pi_M}{\pi_1} = \frac{p_1}{p_M} \gg 1.
\]  
Thus the market maker receives frequent but small profits on metaorders of length one and rare but large losses for metaorders of length $M$.

 {\bf Proposition 3.}~{\it The system of martingale conditions (Eq. \ref{shortterm}) and fair pricing conditions (Eq. \ref{fairPricing}) has solution
\begin{eqnarray}
R^+_t=\frac{1}{t}\frac{p_t}{\sum_{i=t+1}^M p_i}\frac{1-p_1}{\sum_{i=t}^M p_i}R^+_1~~~~~t=2,3,...,M-1\label{solutiona}\\
R^-_t=\frac{{\cal P}_t}{1-{\cal P}_t} R^+_t~~~~~t=1,2,...,M-1
\label{solution2a}
\end{eqnarray}}
 
{\bf Proof.} The solution of Eq. (\ref{solution2a}) is a direct consequence of the martingale conditions (Eq. \ref{shortterm}). The total profit of metaorders of length $N<M$ can be rewritten as (see proof of Proposition 1)
\begin{equation}
N\pi_N=NR^-_N-\sum_{i=1}^{N-1}iR^+_i=N\frac{{\cal P}_N}{1-{\cal P}_N} R^+_N-\sum_{i=1}^{N-1}iR^+_i
\end{equation}
The fair pricing conditions (Eq. \ref{fairPricing}) state that for $1<N<M$ it is $N\pi_N=0$, i.e.
\begin{equation}
R^+_N=\frac{1}{N}\frac{1-{\cal P}_N}{{\cal P}_N}\sum_{i=1}^{N-1}iR^+_i
\label{receqa}
\end{equation}
This is a recursive equation which determines $R^+_N$ once $R^+_1$ is given (note that this equation does not hold for $N=1$ because we do not have fair pricing for metaorders of length one). The solution of this equation is  
\begin{equation}
R^+_t=\frac{1}{t}\frac{1-{\cal P}_t}{{\cal P}_t}\frac{1}{{\cal P}_1{\cal P}_2{\cal P}_3....{\cal P}_{t-1}}R^+_1~~~~~t>1
\label{solreceqa}
\end{equation}
and we prove it by induction. We assume that the solution holds for $N=2,3,...,t-1$ and we prove that it is true for $N=t$. If Eq. (\ref{solreceqa}) holds for $N=2,3,..,t-1$ we can rewrite Eq. (\ref{receqa}) for $N=t$ as
\begin{equation}
R^+_t=\frac{1}{t}\frac{1-{\cal P}_t}{{\cal P}_t}\sum_{i=1}^{t-1}iR^+_i=\frac{1}{t}\frac{1-{\cal P}_t}{{\cal P}_t}\left(1+\sum_{i=2}^{t-1}\frac{1-{\cal P}_i}{{\cal P}_i}\frac{1}{{\cal P}_1{\cal P}_2{\cal P}_3....{\cal P}_{i-1}}\right)R^+_1
\end{equation}
Now by expanding the sum in brackets it is direct to show that
\begin{equation}
\left(1+\sum_{i=2}^{t-1}\frac{1-{\cal P}_i}{{\cal P}_i}\frac{1}{{\cal P}_1{\cal P}_2{\cal P}_3....{\cal P}_{i-1}}\right)=1-\frac{1}{{\cal P}_1}+\frac{1}{{\cal P}_1{\cal P}_2{\cal P}_3....{\cal P}_{t-1}}
\end{equation} 
Since, by definition, ${\cal P}_1=1$ the first two terms in the right hand side cancel and thus one obtains Eq. (\ref{solreceqa}).  This equation is equivalent to Eq. (\ref{solutiona}). In fact
\begin{eqnarray}
R^+_t=\frac{1}{t}\frac{1-{\cal P}_t}{{\cal P}_t}\frac{1}{{\cal P}_1{\cal P}_2{\cal P}_3....{\cal P}_{t-1}}R^+_1=
\frac{1}{t}\frac{p_t}{\sum_{i=t+1}^Mp_i}\frac{1}{\frac{\sum_{i=3}^Mp_i}{\sum_{i=2}^Mp_i}\frac{\sum_{i=4}^Mp_i}{\sum_{i=3}^Mp_i}.....\frac{\sum_{i=t}^Mp_i}{\sum_{i=t-1}^Mp_i}} R^+_1=\nonumber \\
=\frac{1}{t}\frac{p_t}{\sum_{i=t+1}^M p_i}\frac{\sum_{i=2}^Mp_i}{\sum_{i=t}^M p_i}R^+_1=\frac{1}{t}\frac{p_t}{\sum_{i=t+1}^M p_i}\frac{1-p_1}{\sum_{i=t}^M p_i}R^+_1
\end{eqnarray}
i.e. our thesis, Eq. (\ref{solutiona}).

\section{Market impact for other metaorder size distributions: stretched exponential and lognormal}

Changing the metaorder distribution has a dramatic effect on the impact.   While we believe that the Pareto distribution is empirically the correct functional form for metaorder size, to get more insight into the role of $p_N$ we compute the impact for two alternative functional forms.

The first one is the stretched exponential, which can be tuned from thin tailed  to heavy tailed behavior and contains the exponential distribution as a special case.  There is no simple expression for the normalization factor needed for a discrete stretched exponential distribution, so we make a continuous approximation, in which the metaorder size distribution is
\begin{eqnarray}
p_N = \frac{\lambda}{\Gamma(1/\lambda, 1)} e^{-N^\lambda}. 
\end{eqnarray}
The normalization factor $\Gamma(a,z)$ is the incomplete Gamma function.  The shape parameter $\lambda > 0$ specifies whether the distribution decays  faster or slower than an exponential.  ($\lambda > 1$ implies faster decay and $\lambda < 1$ implies slower decay.)   For short data sets, when $\lambda$ is small  this functional form is easily confused with a power law.  

It can be shown that the stretched exponential leads to an immediate impact function that for large $t$ asymptotically  behaves as
\begin{eqnarray}
\mathcal{I}_t \sim \frac{e^{t^\lambda}}{t^{2-\lambda}}.
\end{eqnarray}
This is the product of a power law and an exponential; for large $t$ the exponential dominates.  The permanent impact is
\begin{equation}
I_N \sim \frac{1}{N}\int^N \frac{e^{x^\lambda}}{x^{2-\lambda}}dx \sim\frac{e^{N^\lambda}\lambda-E_{1+1/\lambda}(-N^\lambda)}{N^2 \lambda^2}\sim \frac{e^{N^\lambda}}{N^2\lambda},
\end{equation}
where $E_\nu(z)$ is the exponential integral function and in the last approximation we have used its asymptotic expansion.  The ratio of the permanent to the immediate impact of the last transaction is
\begin{equation}
{I_N \over \mathcal{I}_N} = \frac{1}{\lambda N^\lambda}.
\end{equation}
In contrast to the Pareto metaorder size distribution this is not constant.   Instead the ratio between permanent and immediate impact decreases with size, going to zero in the limit as $N \to \infty$.   

A similar result is obtained for the case in which metatorder size is lognormal distributed. One of the reasons for considering this form is the empirical results reported in Vaglica et al (2008) where it was shown that the power-law behavior of metaorder size comes from the heterogeneity of market participants and metaorder size distribution of individual institutions is better fit by a lognormal distribution.

Let us consider for simplicity a standardized lognormal distribution

\begin{equation}
p_N=\frac{1}{N\sqrt{2\pi}}e^{\log^2(N)/2},
\end{equation} 
where we have again used the continuous approximation. Note that qualitatively similar results are obtained with non standardized lognormal (i.e. with $\mu\ne 0$ and $\sigma\ne 0$), but the expressions are longer and less transparent.

The immediate impact function for large $t$ can be computed and it is
\begin{equation}
\mathcal{I}_t \sim  e^{\log^2(t)/2} \frac{\log(t)}{t}
\end{equation}
As expected, this is an increasing convex function of $t$, at odds with empirical data. The permanent impact is
\begin{equation}
I_N \sim \frac{1}{N}\int^N e^{\log^2(x)/2} \frac{\log(x)}{x} dx=\frac{1}{N} e^{\log^2(N)/2}
\end{equation}
The ratio of the permanent to the immediate impact of the last transaction is
\begin{equation}
{I_N \over \mathcal{I}_N} = \frac{1}{\log N}.
\end{equation}
As for the stretched exponential case the ratio is not constant but decreases with metaroder size $N$. The fixed ratio of permanent and immediate impact is a rather special property of the Pareto distribution. 

\section{Effect of finite $M$ on impact}

As already discussed briefly in Section~\ref{finiteSize}, if the condition $N \ll M$ is violated this has an effect on the impact.  In this section we consider the exact case of a finite support Pareto distribution. We show that when $N \ll M$ we obtain the same results of the previous section and we discuss what happens when $N \approx M$.

We assume that the metaorder size distribution is a truncated Pareto distribution for all $N\le M$, i.e.
\begin{equation}
p_N=\frac{1}{H_M^{(1+\beta)}}\frac{1}{N^{\beta+1}}~~~~~~~~~~~~~N\ge1
\label{paretoTrunc}
\end{equation}
where the normalization constant $H_M^{(1+\beta)}$ is the harmonic number of order $1+\beta$.  
For the truncated Pareto distribution the probability ${\cal P}_t$ that a metaorder of size $t$ will continue is
\begin{equation}
{\cal P}_t=\frac{\zeta(1+\beta,t+1)-\zeta(1+\beta,M+1)}{\zeta(1+\beta,t)-\zeta(1+\beta,1+M)}
\label{calPApprox2}
\end{equation}
where $\zeta(s,a)$ is the generalized Riemann zeta function (also called the Hurwitz zeta function).
For small $t$ the function ${\cal P}_t$ increases meaning that it is more and more likely that the order continues.  In the regime of $ t\ll M$, ${\cal P}_t$ is well approximated by the expression of Eq. (\ref{calPApprox}) for an infinite support Pareto distribution.  However, around $t\simeq M/2$, ${\cal P}_t$ starts to decrease meaning that  it becomes more and more likely that the order is going to stop soon, with a corresponding effect on the impact.

The immediate impact can be easily calculated once the distribution of metaorder size is known by using Eq. (\ref{final}). 
For  truncated Pareto distributed metaorder sizes, $R^+_t$ is (for $t>2$) is
\begin{equation}
R^+_t=\left(\frac{H_M^{(1+\beta)}-1}{(\zeta(1+\beta,t)-\zeta(1+\beta,M+1))(\zeta(1+\beta,t+1)-\zeta(1+\beta,M+1)}\right)\frac{R^+_1}{t^{2+\beta}}
\label{paretosolTrunc}
\end{equation}
For large $t$ but $t\ll M$ it is
\begin{equation}
R^+_t\sim\frac{R^+_1}{t^{2-\beta}}
\end{equation}
which is the same scaling as the infinite support Pareto distribution (see Eq. (\ref{temporaryImpact})).  The same holds true for the permanent impact. We have therefore shown that when  $t\ll M$ the finite support of the metaorder size distribution is irrelevant and we obtain approximately the same results as in Section~\ref{paretoSec}.
 
\begin{figure}[ptb]
\begin{center}
\includegraphics[scale=0.4]{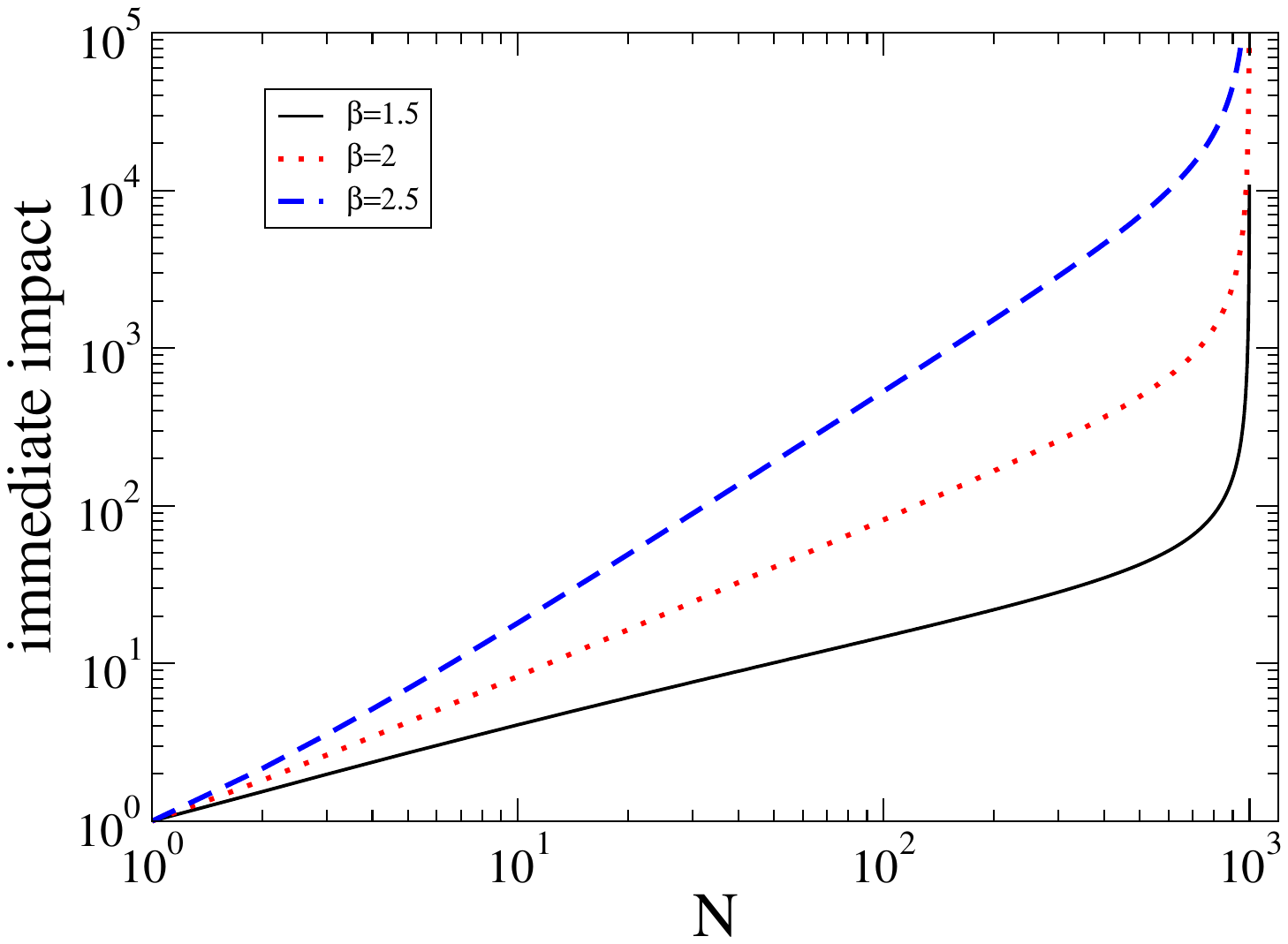}
\end{center}
\caption{Immediate impact when the metaorder size has finite support, i.e when $N$ has a maximum value $M$.  Plot is in log log scale with $M=1000$ and $\beta=1.5$, $2$, and $2.5$.}
\label{figimpact}
\end{figure}

The finite size effects and the role of the finiteness of the support becomes relevant when $t\gtrsim M/2$. 
Figure \ref{figimpact} shows the total impact for $M=1000$ and different values of $\beta$. It is clear that the impact is initially described by a power law, but then it becomes strongly convex when the order length becomes comparable with the maximal length.

\end{document}